\documentclass[aps,preprint,epsfig,rotate]{revtex4}
\usepackage{graphicx}
\usepackage{bm}
\usepackage{epsfig}

\begin{document}
\title{On the absorption of radiation by the negatively charged hydrogen ion. I.
       General theory and construction of the wave functions.}

 \author{Alexei M. Frolov}
 \email[E--mail address: ]{afrolov@uwo.ca}

\affiliation{Department of Applied Mathematics \\
 University of Western Ontario, London, Ontario N6H 5B7, Canada}

\date{\today}

\begin{abstract}

The absorption of infrared and visible radiation from stellar emission spectra by the negatively 
charged hydrogen ions H$^{-}$ is considered. We derive the explicit formulas for the bound-free 
absorption coefficient $a_{\nu}$ for the negatively charged hydrogen ion ${}^{\infty}$H$^{-}$ and 
its isotopes ${}^{1}$H$^{-}$ (protium) and ${}^{2}$H$^{-}$ (deuterium or D$^{-}$). The bound-free 
absorption coefficient $a_{\nu}$ is used to evaluate the actual absorption of infrared and visible 
radiation by the H$^{-}$ ion in photospheres of many cold stars with surface temperatures $T_s \le$ 
8,250 $K$.

\end{abstract}
\maketitle
\newpage

\section{Introduction}

As is well known (see, e.g., \cite{Sob} - \cite{Zir}) the negatively charged
hydrogen ions ${}^{1}$H$^{-}$ (protium) and ${}^{2}$H$^{-}$ (deuterium or
D$^{-}$) are of great interest in Stellar Astrophysics. Indeed, the
negatively charged hydrogen ions substantially determine the absorption of
infrared and visible radiation in photoshperes of all stars with temperatures 
bounded between $T_{max} \approx$ 8,250 $K$ (late A-stars) and $T_{min} 
\approx$ 2,750 $K$ (early M-stars). The main contribution to the light 
absorption by the negatively charged hydrogen ion H$^{-}$ comes from its 
photodetachment
\begin{equation}
 {\rm H}^{-} + h \nu = {\rm H} + e^{-} \label{e0}
\end{equation}
where the notation $h \nu$ designates the incident light quantum, while 
$e^{-}$ means the free electron, or photo-electron, for short. Here and 
below $h$ is the Planck constant. Note that each of the 
negatively charged hydrogen ions, i.e. the ${}^{1}$H$^{-}$, ${}^{2}$H$^{-}$ 
(or D$^{-}$), ${}^{3}$H$^{-}$ (or T$^{-}$) and ${}^{\infty}$H$^{-}$ has 
only one bound (ground) $1^1S-$state with $L = 0$, where $L$ is the angular 
momentum of this two-electron ion (see, e.g., \cite{Fro2005}). The ground
$1^1S-$state is a singlet state, i.e., the total electron spin $S$ in this 
state equals zero, and therefore, $2 S + 1 = 1$. Here and everywhere below 
the notation ${}^{\infty}$H/${}^{\infty}$H$^-$ stands for the model atom/ion 
with the infinitely heavy nucleus. The total non-relativistic energy of the ground 
$1^1S-$state in the ${}^{\infty}$H$^-$ ion is -0.527751 016544 377196 59055(5)
$a.u.$ \cite{Fro2005}, while the total energy of the hydrogen atom with 
the infinitely heavy nucleus (${}^{\infty}$H) is -0.5 $a.u.$ Therefore, the 
binding energy of the ${}^{\infty}$H$^-$ ion is $\chi_1 \approx$ 0.027 751 
$a.u.$ $\approx$ 0.755 143 9 $eV$ and the corresponding ionization frequency 
$\nu_1 = \frac{\chi_1}{h}$ is located in the far infrared part of the 
spectrum. Note that the energies of `visible' light quanta are $\approx$ 2 - 
3 $eV$, while the total energy of the ground state of hydrogen atom is $\approx$
13.605 $eV$. Light quanta with such energies ($\approx$ 13.605 $eV$) correspond 
to the vacuum ultraviolet region. 

The final atomic state which arises after photodetachment of the negatively
charged hydrogen ion, Eq.(\ref{e0}), includes the neutral hydrogen atom H
and free electron $e^{-}$. The kinetic energy of this free electron can, in
principle, be arbitrary. Therefore, it is easy to predict that all radiation
quanta with the frequencies $\nu \ge \nu_1 = \frac{\chi_1}{h}$ can produce
photodetachment of the negatively charged hydrogen ion H$^{-}$. In other 
words, all light quanta with $\nu \ge \nu_1$ can be absorbed during such a 
process. This includes infrared quanta and quanta of visible light.

The absorption of radiation by the negatively charged hydrogen ions has been
investigated in numerous earlier studies (see, e.g., \cite{Chand} -
\cite{FroSm03} and references therein). In general, the light absorption by
an arbitrary atomic system is determined by separate contributions from the
corresponding bound-bound, bound-free and free-free transitions in this
system. As mentioned above the negatively charged hydrogen ion has only one
bound (ground) $1^1S-$state, i.e.  for the H$^{-}$ ion there is no contribution from the 
bound-bound transitions. The bound-free transitions in the H$^{-}$ ion correspond to the 
photodetachment, Eq.(\ref{e0}), while the free-free (electron) transitions formally 
represent the radiation-stimulated electron scattering in the field of the neutral 
hydrogen atom (or inverse bremsstrahlung, for short)
\begin{equation}
 e^{-} + h \nu + {\rm H} = {\rm H} + e^{-} \label{ee0}
\end{equation}
where the kinetic energy of the final electron differs from its incident
energy due to the photon's absorption. If the kinetic energy of the both 
incident photon and electron in Eq.(\ref{ee0}) are small, then the final 
state of the hydrogen atom coincides with its original (ground) state. 
Indeed, the lowest excitation energy of the ground state of the hydrogen
atom is relatively high $E \approx$ 0.375 $a.u.$ $\approx$ 10.2043 $eV$.
Therefore, we can ignore all possible excitations of the central 
hydrogen atom during radiation-stimulated electron scattering in the 
field of the neutral hydrogen atom. Such an approximation has a good 
numerical accuracy, if the kinetic (or thermal) energies of scattered electrons 
$E_e \sim k T$ are significantly smaller than the lowest excitation 
energy of the ground state of the hydrogen atom ($\approx$ 0.375 $a.u.$ 
$\approx$ 10.2043 $eV$). It can be shown that such an assumption is 
always obeyed in all stars in which the absorption of radiation by the 
negatively charged hydrogen ions is important, i.e. it is true for the 
late A-stars and all F, G and K stars. Note that in the early F-stars 
and late A stars the absorption of radiation by the negatively charged 
hydrogen ions is consequently outweighed by the neutral hydrogen 
absorption. In the hot O- and B-stars, with surface temperatures $T_s 
\ge 28,500$ $K$ and $T_s \ge 11,000$ $K$, respectively, the absorption 
of radiation quanta is mainly related with the transitions between 
different bound states in helium and hydrogen atoms. The negatively 
charged H$^{-}$ ion does not exist at such temperatures as a stable, 
i.e. bound, system.

On the other hand, the H$^{-}$ ions play practically no role for the cold M
and N stars with $T_s \le$ 2,700 $K$ where the absorption of radiation by
various molecular species and atoms of metals becomes important.
However, in the late F, G and early K stars the absorption of infrared and
visible radiation by the negatively charged hydrogen ions is maximal. This
includes our Sun which is a star of spectral type G2. The great role of the
H$^{-}$ ions for our Sun was suggested by R. Wild in 1939 (see discussions
and references in \cite{Sob}, \cite{Aller} and \cite{Zir}). An effective 
absorbtion of large amount of infrared solar radiation by the negatively 
charged hydrogen ions is crucial for correct thermal balance of our planet.  

In this study we analyze the bound-free transitions in the negatively charged hydrogen ions. Our main goal is to 
re-consider analytical derivation of formulas which are currently used to calculate the photodetachment 
cross-section of the H$^{-}$ ion. In particular, we want to re-evaluate the validity of some approximations made 
in earlier works where the absorption of radiation by the negatively charged hydrogen ions has been evaluated 
\cite{Chand} - \cite{FroSm03} (see also the references in these papers). We also consider the contribution of 
other channels in the process of photodetachment, Eq.(\ref{e0}). Usually, many of such channels are ignored in 
modern numerical calculations. For instance, a few very papers considered a possibility to form the final hydrogen 
atom in any of its excited (bound) states. The competing $ps-$channel of photodetachment, Eq.(\ref{e0}), was 
completely ignored in earlier studies. Moreover, in many of earlier papers authors considered formation of
the final hydrogen atom in the ground state. In other works, the formation of the final hydrogen atom in a 
number of excited $s-$states was considered, but numerical results were not accurate, since for both the 
incident H$^{-}$ ion and final (`free') electron only very approximate wave functions were used. It should be 
mentioned here taht all Hartree-Fock based methods cannot produce the actual bound state in the H$^{-}$ ion.

Currently, the photodetachment cross-section of the H$^{-}$ ion must be determined to very high accuracy ($\pm$ 0.5 \%). 
The moderate accuracy known from earlier works ($\pm$ 6 \% - 10 \%) is not sufficient to describe effects directly 
related to the climate change and surface temperature variations. In our study we try to improve the accuracy of numerical
evaluations of the photodetachment cross-sections for the H$^{-}$ ion(s). Below, we pay a significant attention to the 
formulas which are used in calculations of the bound-free absorption coefficient of the H$^{-}$ ion. Results of different 
numerical calculations are discussed here very briefly, since many of them are included in our next paper. 

Our original plan also included accurate numerical evaluation of the free-free absorption coefficient for the negatively 
charged hydrogen ions. However, by deriving the corresponding formulas we have found a number of unsolved problems. In 
particular, the idea of inverse bremsstrahlung in the field of atomic polarization potentials is not welcomed by many
authors in Quantum Electrodynamics. Another problem is a substantial energy-dependence of the atomic polarization potential.
Combination of all these facts lead to the conclusion that accurate numerical evaluation of the free-free absorption 
coefficient for the negatively charged hydrogen atom is hardly possible at this time. 

\section{Absorption coefficient of the negatively charged hydrogen ion}

In general, the radiation absorption coefficient $\alpha_{\nu}$ of the negatively charged hydrogen ion defined per unit 
volume is written in the form (see, e.g., \cite{Sob})
\begin{equation}
 \alpha_{\nu} = ( n^{-} k_{\nu} + n_H p_e a_{\nu} ) \Bigl[ 1 -
 \exp\Bigl( -\frac{h \nu}{k T} \Bigr) \Bigr] \label{e1}
\end{equation}
where $n^{-}$ and $n_{\rm H}$ are the spatial densities of the H$^{-}$ ions and neutral hydrogen atoms H, respectively. 
Also, in this formula $p_e = n_e k T$ is the electron pressure which corresponds to the temperature $T =T_e$ and electron 
density $n_e$. The spatial densities $n_H, n^{-}$ and $n_e$ are related to each other by the following equation
\begin{equation}
 \frac{n_H n_e}{n^{-}} = \frac{g_1}{g_{-}} \frac{2 (2 \pi m_e k
 T)^{\frac32}}{h^3} \exp\Bigl( -\frac{\chi_{1}}{k T} \Bigr) \label{e2}
\end{equation}
where $g_1 = 2$ and $g_{-} = 1$ are the statistical weights of the ground states of the neutral hydrogen atom $H$ and 
hydrogen negatively charged ion H$^{-}$, respectively. The notation $\chi_1$ stands for the ionization potential of the 
H$^{-}$ ion. In some works this value is called the detachment potential which is, probably, more correct, since no ion is 
created in the process, Eq.(\ref{e0}). In Eq.(\ref{e2}) and in all formulas below the notations $m_e, h$ and $e$ mean the 
electron mass, Planck constant and electron's electric charge, respectively.

By using Eq.(\ref{e2}) we can express the spatial density of the H$^{-}$ ions (i.e. the $n^{-}$ factor in Eq.(\ref{e1})) 
and reduce Eq.(\ref{e1}) to the form
\begin{equation}
 \alpha_{\nu} = n_{H} p_e \Bigl[ k_{\nu} \frac{h^3}{4 (2 \pi m_e)^{\frac32}
 (k T)^{\frac52}} \exp\Bigl(\frac{\chi_{1}}{k T} \Bigr) + a_{\nu} \Bigr]
 \Bigl[ 1 - \exp\Bigl( -\frac{h \nu}{k T} \Bigr) \Bigr] \label{e3}
\end{equation}
This formula is used for all current calculations of the absorption coefficient $\alpha_{\nu}$ of the negatively charged ion H$^{-}$. In 
real calculations of $\alpha_{\nu}$ it is always assumed that $p_e = 1$ and $n_{H} = 1$, i.e. such calculations are performed for the unit 
electron pressure and per one hydrogen atom in unit volume. In this study the formula, Eq.(\ref{e3}), is also applied for numerical 
calculations of the absorption coefficient $\alpha_{\nu}$ of the H$^{-}$ ion. As it follows from Eq.(\ref{e3}) in order to determine the 
coefficient $\alpha_{\nu}$ one needs to know the bound-free and free-free absorbtion coefficients, i.e. the factors $a_{\nu}$ and $k_{\nu}$, 
respectively. The third unknown value, which can be found in Eq.(\ref{e3}), is the ionization/detachment potential $\chi_{1}$ of the H$^{-}$ 
ion(s). In reality, the ionization potentials $\chi_1$ of all isotope-substituted H$^{-}$ ions are now known to very good numerical accuracy 
(see, e.g., \cite{Fro2005} and Table I below). 

Below we restrict ourselves to the analysis of computational formulas used for the $a_{\nu}$ and $k_{\nu}$ coefficients. Note that the first 
numerical calculations of the absorption coefficient $\alpha_{\nu}$ with the use of Eq.(\ref{e3}) were performed in \cite{Breen}. Formally, 
if the total absorption coefficient $\alpha_{\nu}$ as the function of $r$ in stellar photoshpere is known, then it is possible to determine 
the optical depth at the frequency $\nu$ 
\begin{equation}
 \tau_{\nu}(r) = \int^{+\infty}_{r} \alpha_{\nu}(y) dy
\end{equation}
where $r$ is the actual geometrical distance measured from the top of the solar/stellar photosphere, where $r = \infty$. 

To conclude this Section let us discuss the relations between the absoption coefficients $a_{\nu}$ and $k_{\nu}$ from Eqs.(\ref{e1}) and 
(\ref{e3}) and the corresponding (microscopic) cross-sections used below. Without loss of generality consider such a relation for the 
bound-free coefficient $a_{\nu}$ and photodetachment cross-section $\sigma_{\nu}$ of the H$^{-}$ ion. This relation is obvious: $a_{\nu} = 
\sigma_{\nu} \cdot 1 \cdot N \approx \sigma_{\nu} \frac{\rho}{M_H} N_A$, where $N_A \approx 6.0223 \cdot 10^{23}$ is the Avogadro number, $M_H$ 
is the mass of one mole of hydrogen atoms and $\rho$ is the density of the stellar (hydrogenic) photosphere. 

\section{Photodetachment of the negatively charged hydrogen ion}

In this Section we derive and discuss the explicit formulas which are used for numerical computation of 
the photodetachment cross-section of the H$^{-}$ ion. The photodetachment corresponds to the bound-free 
(optical) transitions, since the final state of one of the two electrons is an unbound state, or, in 
other words, the state from the unbound spectra of the H$^{-}$ ion. For the two-electron H$^{-}$ ion it 
is possible to obtain the closed analytical formula for the photodetachment cross-section, if the 
non-relativistic (or dipole) approximation is used. The derivation of this formula is the main goal of 
this Section. 

Let us assume that the incident (i.e. second) electron was bound to the neutral hydrogen atom with the 
binding energy $\epsilon_{i} = -I$, where $I = \chi_1$ is the ionization potential of the H$^{-}$ ion. The 
incident photon has the momentum ${\bf k}$ and the frequency $\omega$ (or energy $\hbar \omega = \omega$). 
In the final state the emited photo-electron moves as a free particle with the momentum ${\bf p}$ and 
energy $\epsilon_f$. Since ${\bf p}$ is a continuous variable, the photodetachment cross-section is written 
in the form 
\begin{equation}
 d\sigma = 2 \pi \mid M_{i \rightarrow f} \mid^2 \delta(-I + \omega - \epsilon_f) \frac{d^3 p}{(2 \pi)^3} \label{cross}
\end{equation}
where $M_{i \rightarrow f}$ is the transition amplitude (see, Eq.(\ref{Matrix}), below) and the wave function 
of the final state is normalized per one particle in unit volume $V = 1$. All wave functions and expressions 
used in computations of the transition amplitude $M_{i \rightarrow i}$ are assumed to have a proper permutation 
symmetry upon spin-spatial coordinates of the two electrons 1 and 2. This allows one to write the formula, 
Eq.(\ref{cross}), and all formulas below in the one-electron form.

The delta-function in Eq.(\ref{cross}) is excluded by integrating over the momenta ${\bf p}$ of the photo-electron. Indeed, by using 
the formula $d^3{\bf p} = p^2 d\mid {\bf p} \mid do = \epsilon \mid {\bf p} \mid d\epsilon do$ and performing the integration over 
$d\epsilon$ we can remove this delta-function and obtain the following general expression for the cross-section
\begin{equation}
 d\sigma = e^2 \frac{\epsilon \mid {\bf p} \mid}{2 \pi \omega} \mid M_{i \rightarrow f} \mid^2 do \label{dcross}
\end{equation}
The derivation of the analytical formula for the transition amplitude $M_{i \rightarrow f}$ is drastically simplified (see, e.g., 
\cite{AB}) with the use of transverse gauge for the vector potential ${\bf A}$, i.e. $div{\bf A} = 0$. In this case the transition 
amplitude $M_{i \rightarrow f}$ is written as a scalar product of the vector potential ${\bf A}$ and transition current ${\bf j}_{i 
\rightarrow f}$, i.e. $M_{i \rightarrow f} = - e {\bf A} \cdot {\bf j}_{i \rightarrow f}$. The expressions for the vector potential
${\bf A}$ (in the case when $div {\bf A} = 0$) and for the transition current ${\bf j}_{i \rightarrow f}$ can be found in standard 
QED books (see, e.g., \cite{AB}). With these expressions we can write for the transition amplitude
\begin{eqnarray}
 M_{i \rightarrow f} = - e {\bf A} \cdot {\bf j}_{i \rightarrow f} = - e \sqrt{\frac{2 \pi}{\omega}} \int \int \psi^{*}_f(1,2) 
 ({\bf e} \cdot {\bf \alpha}_1) \exp(\imath {\bf k}_{f} \cdot {\bf r}_1) \psi_i(1,2) d^3{\bf r}_1 d^3{\bf r}_2 \label{Matrix}
\end{eqnarray}
where $\psi_i(1,2)$ and $\psi_f(1,2)$ are the wave functions of the incident and final atomic systems. These two functions are assumed 
to be properly symmetrized in respect to all possible electron-electron permutations and each of these functions has a unit norm. Also, 
in this equation $\alpha_1$ are the three Dirac matrixes of the first electron, ${\bf e}$ and ${\bf k}_{f}$ are two vectors which describe 
polarization and direction of propagation of the incident photon, respectively.  

In this study all electrons are considered as non-relativistic particles, while the energy of the incident photon $\hbar \omega (= \omega)$ 
is assumed to be larger than $I$, but substantially smaller than the energy of electron at rest, i.e. $\hbar \omega \ll m_e c^2$. This 
means that the final velocity of the photoelectron is small, i.e. we are dealing with the non-relativistic problem. Therefore, the Dirac's 
$\alpha$ matrixes in Eq.(\ref{Matrix}) can be replaced by the corresponding components of the velocity ${\bf v}_1 = -\frac{\imath}{m_e} 
\nabla_1$. Moreover, in the non-relativistic approximation we can replace the factor $\exp(\imath {\bf k}_f \cdot {\bf r}_1)$ by uinty. This 
gives the following formula for the photodetachment cross-section
\begin{eqnarray}
 d\sigma_{\nu} &=& \frac{e^2 \mid {\bf p} \mid}{m_e \pi \omega} \Bigl| \Bigl\{\int \int \Psi^{*}_{fi}({\bf r}_1, {\bf r}_2) 
 \Bigl[ {\bf e} \cdot \Bigl(\frac{\partial}{\partial {\bf r}_1} \Bigr)\Bigr] \Psi_{{\rm H^{-}}}({\bf r}_1, {\bf r}_2) d^3{\bf r}_1 d^3{\bf r}_2 
 \Bigr\} \Bigr|^2 do \label{knu} \\
 &=& \frac{e^2 \mid {\bf p} \mid}{2 m_e \pi \omega} \Bigl| \Bigl\{\int \int 
 \Psi^{*}_{fi}({\bf r}_1, {\bf r}_2) \Bigl[ {\bf e} \cdot \Bigl(\frac{\partial}{\partial {\bf r}_1} + \frac{\partial}{\partial 
 {\bf r}_2}\Bigr) \Bigr] \Psi_{{\rm H^{-}}}({\bf r}_1, {\bf r}_2) d^3{\bf r}_1 d^3{\bf r}_2 \Bigr\} \Bigr|^2 do \nonumber
\end{eqnarray}
where ${\bf e}$ is the vector which represents the polarization of the incident photon. In the case of photodetachment, Eq.(\ref{e0}), the 
notation  $\Psi_{fi}({\bf r}_1, {\bf r}_2)$ in Eq.(\ref{knu}) stands for the wave function of the final state (after photodetachment), while 
$\Psi_{{\rm H^{-}}}({\bf r}_1, {\bf r}_2)$ is the wave function of the ground $1^1S-$state of the hydrogen ion. In atomic units the last formula
from Eq.(\ref{knu}) takes the form 
\begin{eqnarray}
 d\sigma_{\nu} &=& \frac{c}{v_a} \alpha^2 a^2_0 \frac{\mid {\bf p} \mid}{2 \pi \omega} \Bigl| \Bigl\{\int \int 
 \Psi^{*}_{fi}({\bf r}_1, {\bf r}_2) \Bigl[ {\bf e} \cdot \Bigl(\frac{\partial}{\partial {\bf r}_1} + \frac{\partial}{\partial 
 {\bf r}_2}\Bigr) \Bigr] \Psi_{{\rm H^{-}}}({\bf r}_1, {\bf r}_2) d^3{\bf r}_1 d^3{\bf r}_2 \Bigr\} \Bigr|^2 do \label{knu1} \\
 &=& \alpha a^2_0 \frac{\mid {\bf p} \mid}{2 \pi \omega} \Bigl| \Bigl\{\int \int 
 \Psi^{*}_{fi}({\bf r}_1, {\bf r}_2) \Bigl[ {\bf e} \cdot \Bigl(\frac{\partial}{\partial {\bf r}_1} + \frac{\partial}{\partial 
 {\bf r}_2}\Bigr) \Bigr] \Psi_{{\rm H^{-}}}({\bf r}_1, {\bf r}_2) d^3{\bf r}_1 d^3{\bf r}_2 \Bigr\} \Bigr|^2 do \nonumber
\end{eqnarray}
where $c$ is the speed of light in vacuum, $a_0$ is the Bohr radius, $v_a = \frac{e^2}{\hbar}$ is the atomic velocity (the dimensionless ratio 
$\frac{c}{a_0} = \alpha^{-1}$) and all other values must be expressed in atomic units. The formula, Eq.(\ref{knu}), is used in numerical calculations 
of the photodetachment cross-section of the H$^{-}$ ion. Note that the formula, Eq.(\ref{knu}), is written in the two forms: one-electron and 
two-electron. The one-electron form is used in calculations, while two-electron form is needed to show the explicit invariance of the final 
expressions upon all possible electron-electron permutations. To expain all essential details of such calculations we need to discuss the expressions 
which are used to approximate the wave function of the bound $1^1S-$state in the H$^{-}$ ion and wave functions of the final states, i.e. states 
which arise after photodetachment. 

\section{Wave functions of the negatively charged hydrogen ions}

The wave functions of the ground states in the hydrogen ions (${}^{\infty}$H$^{-}$, ${}^{3}$H$^{-}$ (or T$^{-}$), ${}^{2}$H$^{-}$ (or D$^{-}$) and 
${}^{1}$H$^{-}$) are approximated with the use of different variational expansions. One of the best such expansions is the exponential variational 
expansion in relative/perimetric coordinates coordinates. For the ground $1^1S-$state of the H$^{-}$ ion it takes the form
\begin{eqnarray}
 \Psi_{{\rm H^{-}}} &=& \Psi_{{\rm H^{-}}}({\bf r}_1, {\bf r}_2) = \frac{1}{\sqrt{2}} (1 + \hat{P}_{12}) \sum^{N}_{i=1} C_i f_i \label{exp} \\
 &=& \frac{1}{\sqrt{2}} \sum^{N}_{i=1} C_i \Bigl[ \exp(-\alpha_i r_{32} -\beta_i r_{31} - \gamma_i r_{21}) +  
                                                   \exp(-\beta_i r_{32} -\alpha_i r_{31} - \gamma_i r_{21}) \Bigr] \nonumber
\end{eqnarray}
where $\hat{P}_{12}$ is the permutation of the two identical particles 1 and 2 (electrons), the subsript 3 designates the hydrogen nucleus. In this 
notation $r_{21} = \mid {\bf r}_2 - {\bf r}_1 \mid$ is the electron-electron (or correlation) coordinate (scalar) and $r_{3i} = \mid {\bf r}_3 - 
{\bf r}_i \mid$ are the corresponding electron-nuclear coordinates ($i = 1, 2$). Also in this formula $C_i$ ($i$ = 1, $\ldots, N$) are the linear 
variational coefficients of the approximate (or trial) wave function. The $3 N-$parameters $\alpha_i, \beta_i, \gamma_i$ ($i$ = 1, $\ldots, N$) are 
the non-linear parameters (or varied parameters) of the trial wave function, Eq.(\ref{exp}). As follows from the results of our earlier studies (see, 
e.g., \cite{Fro2005}, \cite{Fro98}) the exponential variational expansion, Eq.(\ref{exp}), provides very high numerical accuracy for any bound state 
in arbitrary Coulomb three-body systems, i.e. for systems with arbitrary particle masses and electric charges. For the H$^{-}$ ions such an accuracy 
is much higher than accuracy which any other variational expansion may provide, if the same number of basis functions $N$ is used. Note also that from 
Eq.(\ref{exp}) one finds the following formula
\begin{eqnarray}
 {\bf e} \cdot \Bigl(\frac{\partial}{\partial {\bf r}_1} + \frac{\partial}{\partial {\bf r}_2}\Bigr) \Psi_{{\rm H^{-}}} =
 \Bigl({\bf e} \cdot \frac{{\bf r_{1}}}{r_{1}}\Bigr) \sum^{N}_{i=1} C_i \alpha_i f_i + \Bigl({\bf e} \cdot \frac{{\bf
 r_{2}}}{r_{2}}\Bigr) \sum^{N}_{i=1} C_i \beta_i f_i \\
  = \Bigl({\bf e} \cdot {\bf n}_{1}\Bigr) \Bigl[\frac{\partial}{\partial r_{1}} \Psi_{{\rm H^{-}}}\Bigr]
 + \Bigl({\bf e} \cdot {\bf n}_{2} \Bigr) \Bigl[\frac{\partial}{\partial r_{2}} \Psi_{{\rm H^{-}}}\Bigr] \nonumber
\end{eqnarray}
where ${\bf n}_i = \frac{{\bf r}_{3i}}{r_{3i}}$ is the unit vector of the $i$-th electron and $i$ = 1, 2. Note that in this expression all terms which contain 
the $\gamma_{i}$ parameters cancell each other, since we always have $\frac{\partial r_{21}}{\partial {\bf r}_{1}} = -\frac{\partial r_{21}}{\partial 
{\bf r}_{2}}$.

The presence of a large number of the varied non-linear parameters $\alpha_i, \beta_i, \gamma_i$ in Eq.(\ref{exp}) allows one to construct very compact, 
extremely flexible and highly accurate approximations to the actual wave functions of arbitrary three-body systems, including the negatively charged 
hydrogen ions. The results of our calculations obtained with the use of the exponential expansion can be found in Table I. This Table contains only the 
total energies (in atomic units) and ionization potentials $\chi_1$ (in $eV$). In addition to the values shown in Table I we have determined a large number 
of expectation values which can be used to evaluate various bound state properties and probabilities of dfifferent processes/reactions in these ions. Some 
of these expectation values are shown in Tables II and III. The meaning of the notations used to designate some bound state properties is explained in
\cite{Fro98}. Note that the total non-relativistic energies and other expectation values obtained in our calculations are the most accurate values ever 
obtained for isotope substituted hydrogen ions: ${}^{3}$H$^{-}$ (or T$^{-}$), ${}^{2}$H$^{-}$ (or D$^{-}$) and ${}^{1}$H$^{-}$. These expectation values
indicate the overall quality of our wave functions used in calculations of the photodetachment cross-section of the H$^{-}$ ion(s). A well known test for
bound state wave functions of Coulomb few-body systems is numerical coincidence of the predicted and computed cusp values (see, e.g., \cite{Fro98}). Another 
effective test of the quality of bound state wave functions is the accurate evaluation of the lowest-order QED correction for the hydrogen ions (see 
Appendix I).    

It sould be mentioned that in our calculations of the ${}^{3}$H$^{-}, {}^{2}$H$^{-}$ and ${}^{1}$H$^{-}$ hydrogen ions we have used the following updated 
values for the nuclear masses of the tritium, deuterium and protium (in $MeV/c^{2}$)
\begin{eqnarray}
 m_{e} = 0.510998910 \; \; \; , \; \; \; m_{p} = 938.272046 \\
 m_{d} = 1875.612859 \; \; \; , \; \; \; m_{t} = 2808.290906 \nonumber
\end{eqnarray}
In our calculations these masses are considered as exact. The masses of these three hydrogenic nuclei have recently been determined in high-energy experiments 
to better accuracy than they were known in the middle of 1990's. Usually, these masses are expressed in special high-energy mass units ($MeV/c^{2}$). Numerical 
value of the electron mass $m_e$ can be used to re-calculate these masses to atomic units. Highly accurate computations of the ground states in the 
${}^{3}$H$^{-}$ (or T$^{-}$), ${}^{2}$H$^{-}$ (or D$^{-}$) and ${}^{1}$H$^{-}$ ions with these nuclear masses have never been performed.

\section{The final state wave functions}

After photodetachment of the H$^{-}$ ion, Eq.(\ref{e0}), in the final state we have the neutral hydrogen atom in one of its bound states and `free' electron which 
moves in the field of the hydrogen atom with the kinetic energy $E_e =  \frac{\hbar^2 k^2}{2 m_e}$, where $\hbar = \frac{h}{2 \pi}$ is the reduced Planck constant,
$k$ is the wave number and $m_e$ is the electron mass. The corresponding wave function of the final state is represented as a product of the bound state wave 
function of the hydrogen atom and the wave function of the free electron. The wave function of the bound $(n \ell m)-$state of the hydrogen atom is written in the 
form (see, e.g., \cite{LLQ}) $\Phi_{n \ell m}(r, \Theta, \phi) = R_{n \ell}(Q, r) Y_{\ell m}(\Theta, \phi)$, where $Y_{\ell m}(\Theta, \phi) = Y_{\ell m}({\bf n})$ 
is a spherical harmonic and $R_{n \ell}(Q, r)$ is the one-electron radial function. The radial function $R_{n \ell}(Q, r)$ takes the form (see, e.g., \cite{LLQ})
\begin{eqnarray}
  R_{n \ell}(Q,r) = \frac{1}{r n} \sqrt{\frac{Q (n - \ell - 1)!}{(n +
  \ell)!}} \Bigl[ \frac{2 Q r}{n} \Bigr]^{\ell + 1} \Bigl\{ 
  \sum^{n-\ell-1}_{k=0}
  \frac{(-1)^k}{k!}
  \left(
  \begin{array}{c}
   n + \ell \\
   2 \ell + k + 1
  \end{array}
  \right)
  \Bigl[ \frac{2 Q r}{n} \Bigr]^{k} \Bigr\} \times \nonumber \\
 \exp\Bigl(-\frac{Q r}{n}\Bigr) \label{hydrogen}
\end{eqnarray}
where $Q$ (= 1) is the nuclear charge, while $n$ and $\ell$ are the quantum numbers of this bound state. Note that all radial functions defined by Eq.(\ref{hydrogen}) 
have unit norms. The six following functions of the hydrogen atom are very improtant for calculations of the photodetachment cross-sections:
\begin{eqnarray}
  R_{10}(r) = 2 \exp(- r) \; \; \; , \; \; \; R_{20}(r) = \frac{1}{\sqrt{2}} \exp(- \frac{r}{2}) \Bigl( 1 - \frac{r}{2} \Bigr) \; \; \; , \; \; \; 
  R_{21}(r) = \frac{1}{2 \sqrt{6}} r \exp(- \frac{r}{2}) \label{hydr} \\
  R_{30}(r) = \frac{2}{3 \sqrt{3}}  \Bigl( 1 - \frac{2 r}{3} + \frac{2 r^2}{27} \Bigr) \exp(- \frac{r}{3}) \; \; \; , \; \; \;  
  R_{31}(r) = \frac{8}{27 \sqrt{6}}  \Bigl( 1 - \frac{r}{6} \Bigr) \exp(- \frac{r}{3}) \nonumber
\end{eqnarray}
The final states with these hydrogen wave functions provide largest numerical contributions in the photodetachment cross-section.

The wave function of the free electron is represented as the sum of the plane wave $\phi({\bf r}) = \exp(\imath {\bf k} \cdot {\bf r})$ and converging spherical waves 
which are usually designated as $\phi^{(-)}_{{\bf p}}({\bf r})$ \cite{LLQ}. In the non-relativistic dipole approximation used here the following selection rule is 
applied: transitions from the incident $S(L = 0)-$state of the H$^{-}$ ion can only be into the final $P(L = 1)-$state of the two-body (final) system: the H-atom plus 
free electron. It follows from here that in the final states we can detect the H-atom in one of its $s(\ell = 0)-$states and free electron which moves away in the 
$p$-wave. This is the $sp-$channel of the H$^{-}$ ion photodetachment which produces the largest contribution into photodetachment cross-section of the H$^{-}$ ion. 
Another possibility is the formation of the final H-atom in one of its three $p(\ell = 1)-$states and free electron which moves away in the $s-$wave. This represents 
the $ps-$channel in the photodetachment of the H$^{-}$ ion. 

The wave function of the final state for the $sp-$channel takes the form
\begin{eqnarray}
 \psi_f(1,2) = R_{n 0}(Q = 1, r_{31}) \frac{3 \imath}{2 p_e} ({\bf n}_e \cdot {\bf n}_{2}) \Phi_{p 1}(r_{32}) = \frac{3 \imath}{2 p_e} ({\bf n}_e \cdot 
 {\bf n}_2) R_{n 0}(r_{31}) \Phi_{p_e 1}(r_{32}) \label{hfinal}
\end{eqnarray}
where ${\bf n}_2 = \frac{{\bf r}_{32}}{r_{32}}$ is the unit vector of the second (unbound) electron and ${\bf n}_e = \frac{{\bf p}_e}{p_e}$ is the direction of the 
momentum of the outgoing photo-electron ${\bf p}_e$, i.e. the second electron when it becomes unbound after photodetachment. The notation $R_{n \ell=0}(Q = 1, r_{31}) =  
R_{n 0}(r_{31})$ stands for the radial function of the bound state of the hydrogen atom with quantum numbers $n \ge 1$ and $\ell = 0$. In this equation the notation 
$\Phi_{p_e 1}(r_{32})$ stands for the wave function of the electron which becomes `free' after photodetachment. Such a wave function represents an electron which moves in 
the field of the neutral hydrogen atom. This electron has an angular momentum $\ell = 1$ and absolute momentum $p_e = \mid {\bf p}_e \mid$.

The wave function of the final state for the $ps-$channel can be written in the following form
\begin{eqnarray}
 \psi_f(1,2) = R_{n \ell=1}(Q = 1, r_{31}) Y_{1m}({\bf n}_1) \frac{1}{2 p_e} \Phi_{p \ell=0}(r_{32}) = \frac{1}{2 p_e} {\bf n}_{1m} \cdot R_{n 1}(r_{31}) \Phi_{p \ell = 
 0}(r_{32}) \label{arl}
\end{eqnarray}
where $R_{n 1}(Q = 1, r_{31}) = R_{n 1}(r_{31})$ is the radial function of the bound $\mid n 1 \rangle$-state of the hydrogen atom with the principal quantum number $n \ge 
2$ and and angular quantum number $\ell = 1$. Also, in Eq.(\ref{arl}) the notation $Y_{1m}({\bf x})$ stands for the corresponding spherical harmonics \cite{LLQ}, while 
the vector ${\bf n}_{1} = \frac{{\bf r}_{31}}{r_{31}}$ has the following components: $({\bf n}_{1})_{+1} = - \frac{1}{\sqrt{2}} (n_x + \imath n_y), ({\bf n}_{1})_{0} = n_z, 
({\bf n}_{1})_{+1} = - \frac{1}{\sqrt{2}} (n_x - \imath n_y)$, where $n_x, n_y, n_z$ are the three Cartesian components of the unit vector ${\bf n}_1$. Note that all 
calculations for the $ps-$channel in the photodetachment of the H$^{-}$ ion are significantly more complicated than for the $sp-$channel. One obvious complication can 
directly be seen from Eq.(\ref{arl}) where the final wave function is a vector. 

For the $sp-$ and $ps-$channels mentioned above the radial functions of the final states of the hydrogen atom can be designated as the $\mid n 0 \rangle$ and $\mid n 1 
\rangle$ states, respectively. Here the notations $n$ and $\ell$ ($n \ge \ell + 1$) designate the principal quantum number $n$ and angular number $\ell$ of the hydrogen 
atom. To designate the radial hydrogen functions we do not need to mention the magnetic number(s) $m$, where $m = -\ell, -\ell + 1, \ldots, \ell - 1, \ell$. In the case 
of the photodetachment of the hydrogen negative ion H$^{-}$ we always have $\ell = 1$ and $\ell = 0$. It is also clear that the largest contribution to the photodetachment 
of the H$^{-}$ ion comes from that final state in which the final hydrogen atom is in its ground $1^2S-$state with $\ell = 0$. This corresponds to the $sp-$channel. In 
this channel the final hydrogen atoms can also be formed  in the excited $2s(\ell = 0)-, 3s(\ell = 0)-, 4s(\ell = 0)-$states. All these channels also contribute to the 
photodetachment of the H$^{-}$ ion, but their contributions are much smaller than the contribution of the photodetachment channel in which the final hydrogen atom is 
formed in the ground state. Analogous situation can be found for the $ps-$channel where the contribution from the $2p(\ell = 1)-$state substantially exceeds contributions 
from other $np(\ell = 1)-$states in hydrogen atoms (here $n \ge 3$). In general, the formation of the final hydrogen atoms in the highly excited $s(\ell = 0)-$states and 
$p(\ell = 1)-$states with $n \ge 4$ is very unlikely, since the corresponding probabilities are very small. 

It is also clear $a$ $priori$ that the overall contribution of the $sp-$channel in the photodetachment of the H$^{-}$ ion is substantially larger than analogous 
contribution from the $ps-$channel. This directly follows from very large excitation energy of the $2p(\ell = 1)-$states in the hydrogen atom $\Delta E \approx$ 10.2043 
$eV$. Another reason for this is almost exact orthogonality of the wave functions of the $2p-$state of the hydrogen atom H and the ground state of the H$^{-}$ ion, 
respectively. In reality, the overall contribution of the $ps-$channel in the photodetachemnt cross-section of the H$^{-}$ ion is $\approx$ 15 - 25 times smaller than 
analogous contribution of the $sp-$channel. In general, it is comparable with the contributions from the processes which include electric quadrupole and magnetic dipole 
transitions, i.e. processes related with absorption of higher multipole radiation. For these reasons, below in this study we shall consider only the `traditional' 
$sp-$channel of the photodetachment. Note, however, that the $ps-$channel of the photodetachment is of some restricted theoretical interest, since in this case one finds 
no direct correlation between the directions of propagation of the incident photon and final electron. Furthermore, by transforming the integral, Eq.(\ref{knu}), where 
the final wave function is chosen in the form, Eq.(\ref{arl}), one finds that the emission of the photo-electrons in the outgoing $s-$wave(s) is produced during 
the absorbtion of the circularly polarized photons only.  

For stellar astrophysics we can predict that in the cold stars the $ps-$channel does not contribute into the photodetachment of the H$^{-}$ ion. However, the contribution 
of this $ps-$channel rapidly increases with the temperature due to Maxwellian (thermal) distribution of the H$^{-}$ ions upon temperatures in stellar photospheres. It 
follows from here that for stars with higher surface temperatures the $ps-$channel must play a larger role in the photodetachment of the H$^{-}$ ions. In other words, there 
is a fundamental difference between the photodetachment of the H$^{-}$ ions in photospheres of the late A-stars and in photoshperes of earlier K-stars. Note also that in 
our non-relativistic dipole approximation the rotationally excited channels with $\ell \ge 2$ do not contribute in the total photodetachment cross-section.

\section{Formulas for the matrix elements}

For the $sp-$channel the final state wave function is represented in the form of Eq.(\ref{hfinal}) and the formula, Eq.(\ref{knu}), is reduced to the form
\begin{equation}
 \mid \Bigl({\bf e} \cdot \frac{{\bf r_{31}}}{r_{31}}\Bigr) \Phi_{1} + \Bigl({\bf e} \cdot \frac{{\bf r_{32}}}{r_{32}}\Bigr) \Phi_{2} \mid^2 \label{ee1}
\end{equation}
where $\Phi_{1}$ and $\Phi_{2}$ are the two scalar expressions. In stellar photospheres all hydrogen ions are free-oriented. This means that the expression from 
Eq.(\ref{ee1}) must be averaged over possible directions of the unit vectors ${\bf e}_1$ and ${\bf e}_2$ which describe the polarization of the incident photon. In the 
incident light wave we always have ${\bf e}_{1} \perp {\bf e}_{2} \perp {\bf k}_f$, where the unit vector ${\bf k}_f$ corresponds to the direction of propagation of this 
light wave, or the corresponding light quantum (photon). An arbitrary scalar expression can be averaged over all possible directions of the unit vectors ${\bf e}_1$ and 
${\bf e}_2$ with the use of the following formulas
\begin{equation}
 \overline{e_i e_k} = \frac12 (\delta_{ik} - (k_f)_{i} (k_f)_{k})
\end{equation}
and
\begin{equation}
 \overline{({\bf a} \cdot {\bf e}) ({\bf b} \cdot {\bf e})} = \frac12 [({\bf a} \cdot {\bf b}) - ({\bf a} \cdot {\bf k}_f) ({\bf b} \cdot
 {\bf k}_f)] = \frac12 ({\bf a} \times {\bf k}_f) ({\bf b} \times {\bf k}_f) \label{aver}
\end{equation}
where ${\bf a}$ and ${\bf b}$ are the two arbitrary vectors and the notation ${\bf x} \times {\bf y}$ stands for the vector product of the two vectors ${\bf x}$ and 
${\bf y}$. Note that the final formula includes only the unit vector ${\bf k}_f$ which describes the direction of photon propagation. If the direction of the propagation 
of the final photon is not known, then the last formula must also be averaged over all possible directions of the unit vector ${\bf k}_f$. This leads to the following 
formula
\begin{equation}
 \overline{({\bf a} \times {\bf k}_f) ({\bf b} \times {\bf k}_f)} = \frac{8 \pi}{3} ({\bf a} \cdot {\bf b}) 
\end{equation}

Now, in the case of the photodetachment of the H$^{-}$ ion we can write for the following expression for the transition amplitude $M_{i \rightarrow f}$:
\begin{eqnarray}
 & & M_{i \rightarrow f} = \frac{2 \pi}{\sqrt{2} p_e} ({\bf n}_e \cdot {\bf e}) \sum^{N}_{j=1} C_j \Bigl[\alpha_j \int_{0}^{+\infty} 
 \int_{0}^{+\infty} \int^{r_{31} + r_{32}}_{\mid r_{31} - r_{32} \mid} R_{n0}(r_{31}) \Phi_{\ell = 1}(p_e; r_{32}) \times \nonumber \\
 & & \exp(-\alpha_j r_{32} -\beta_j r_{31} - \gamma_j r_{21}) r_{32} r_{31} r_{21} dr_{32} dr_{31} dr_{21} + \beta_j \times \label{int0} \\
  & & \int_{0}^{+\infty} \int_{0}^{+\infty} \int^{r_{31} + r_{32}}_{\mid r_{31} - r_{32} \mid} R_{n0}(r_{31}) \Phi_{\ell = 1}(p_e; r_{32})
 \exp(-\beta_j r_{32} -\alpha_j r_{31} - \gamma_j r_{21}) r_{32} r_{31} r_{21} dr_{32} dr_{31} dr_{21} \Bigr] \nonumber
\end{eqnarray}
where the notation $\Phi_{\ell = 1}(p_e; r_2)$ designates the wave function of the `free' electron which moves in the field of the neutral hydrogen atom.  
As one can see from this formula numerical calculation of the transition amplitude $M_{i \rightarrow f}$ is reduced to the calculation of the three-body integrals in 
relative coordinates $r_{32}, r_{31}, r_{21}$. This problem is discussed in detail in the next Section. Here we want to note that the photodetachment cross-section for 
the $sp-$channel always contains the factor $\mid M_{i \rightarrow f} \mid^2$ which is proportional to the factor $({\bf n}_e \cdot {\bf e})^2$. By applying the 
formula, Eq.(\ref{aver}), one finds that in the dipole approximation the photodetachment cross-section in such cases is always proportional to the factor $({\bf n}_e 
\times {\bf k}_f)^2 = \sin^2 \Theta$, where $\Theta$ is the angle between the directions of propagation of the incident photon ${\bf k}_f$ and final electron 
${\bf n}_e$ (or photo-electron). On the other hand, the presence of this factor in the formula for the photodetachment cross-section indicates clearly that the dipole 
approximation is valid.    

\section{Three-body integrals in relative coordinates}

As we have shown above the problem of analytical and numerical computation of matrix elements arising in Eq.(\ref{knu}) and in Eq.(\ref{int0}) with the trial wave 
function defined in Eq.(\ref{exp}) is reduced to the computation of the following three-body integrals with the Bessel functions $j_{0}(k r_2)$ 
and $j_{1}(k r_2)$ 
\begin{eqnarray}
 \int_{0}^{+\infty} \int_{0}^{+\infty} \int^{r_{31} + r_{32}}_{\mid r_{31} - r_{32} \mid} \phi_{n\ell}(r_{31}) j_{q}(k r_{32}) \exp(-\alpha r_{32} -\beta r_{31} - 
 \gamma r_{21}) r_{32} r_{31} r_{21} dr_{32} dr_{31} dr_{21} \label{int1}
\end{eqnarray}
where $q$ = 0, 1 and $r_{21} = \mid {\bf r}_2 - {\bf r}_1 \mid$ is the electron-electron coordinate which varies between the following limits $\mid r_{32} - r_{31} \mid 
\le r_{21} \le r_{32} + r_{31}$. In Eq.(\ref{int1}) the notation $\phi_{n\ell}(r_{31})$ stands for the unit norm wave functions $R_{n0}(r_{31})$ and $R_{n1}(r_{31})$ of 
the bound states of the hydrogen atom, Eqs.(\ref{hydrogen}) and (\ref{hydr}). All other notations in Eq.(\ref{int1}) are defined above in Eq.(\ref{exp}). The integrals 
defined by Eq.(\ref{int1}) are computed analytically in perimetric coordinates $u_1, u_2, u_3$, where $u_1 = \frac12 (r_{31} + r_{21} - r_{32}), u_2 = \frac12 (r_{32} + r_{21} - 
r_{31})$ and $u_3 = \frac12 (r_{31} + r_{32} - r_{21})$. The three perimetric coordinates $u_1, u_2, u_3$ are independent of each other and each of them varies between 
0 and +$\infty$.

Let us derive the closed analytical expression for the integrals, Eq.(\ref{int1}) with $q = 0$ and $q = 1$. First, we need to obtain the general formula for the closely 
related auxiliary three-body integral $\Gamma_{n,k,l}(\alpha,\beta,\gamma)$ which is written in the form
\begin{eqnarray}
 \Gamma_{n,k,l}(\alpha,\beta,\gamma) =
 \int_{0}^{+\infty} \int_{0}^{+\infty} \int^{r_{31} + r_{32}}_{\mid r_{31} - r_{32} \mid} r^{k}_{32} r^{l}_{31} r^{m}_{21} \exp(-\alpha r_{32} -\beta r_{31} - \gamma r_{21}) 
  r_{32} r_{31} \nonumber \\
  r_{21} dr_{32} dr_{31} dr_{21} \label{int2}
\end{eqnarray}
where $k \ge 0, l \ge 0, n \ge 0$ and $\alpha + \beta > 0, \alpha + \gamma > 0$ and $\beta + \gamma > 0$ (see below). Analytical computation of this integral has extensively been 
explained in a number of our erlier works. Corresppondingly, below we restrict ourselves only to a few following remarks. In perimetric coordinates the integral, Eq.(\ref{int2}), 
takes the form
\begin{eqnarray}
 \Gamma_{n,k,l}(\alpha,\beta,\gamma) = 2 \int_0^{\infty} \int_0^{\infty} \int_0^{\infty} exp\Bigl[-(\alpha + \beta) u_1 -(\alpha + \gamma) u_2
 -(\beta + \gamma) u_3\Bigr] (u_1 + u_2)^n \nonumber \\
 (u_1 + u_3)^m (u_2 + u_3)^l du_1 du_2 du_3 \label{int3}
\end{eqnarray}
where we took into account the fact that the Jacobian of transformation from the relative $(r_{32}, r_{31}. r_{21})$ to perimetric coordinates $(u_1, u_2, u_3)$ equals 2. 
The integration over three independent perimetric coordinates $u_i$ ($0 \le u_i < \infty$) in Eq.(\ref{int3}) is trivial and explicit formula for the 
$\Gamma_{n,k,l}(\alpha,\beta,\gamma)$ integral is written in the form
\begin{eqnarray}
 & &\Gamma_{k;l;n}(\alpha, \beta, \gamma) = 2 \sum^{k}_{k_1=0} \sum^{l}_{l_1=0} 
 \sum^{n}_{n_1=0} C^{k}_{k_1} C^{l}_{l_1} C^{n}_{n_1} 
 \frac{(l-l_1+k_1)!}{(\alpha + \beta)^{l-l_1+k_1+1}}
 \frac{(k-k_1+n_1)!}{(\alpha + \gamma)^{k-k_1+n_1+1}}
 \frac{(n-n_1+l_1)!}{(\beta + \gamma)^{n-n_1+l_1+1}} \nonumber \\
 &=& 2 \cdot k! \cdot l! \cdot n! \sum^{k}_{k_1=0} \sum^{l}_{l_1=0} \sum^{n}_{n_1=0} 
 \frac{C^{k_1}_{n-n_1+k_1} C^{l_1}_{k-k_1+l_1} C^{n_1}_{l-l_1+n_1}}{(\alpha + 
 \beta)^{l-l_1+k_1+1} (\alpha + \gamma)^{k-k_1+n_1+1} (\beta + \gamma)^{n-n_1+l_1+1}} 
 \label{int4}
\end{eqnarray}
where $C^{m}_{k}$ are the binomial coefficients (= number of combinations from $k$ by $m$). In many application all three integer numbers $k, l, m$ are relatively small 
($\le 3$) and this simplifies numerical applications of the formula, Eq.(\ref{int4}).

The parameters $\alpha, \beta$ and $\gamma$ in Eqs.(\ref{int1}) - (\ref{int4}) can be arbitrary real and/or complex numbers. Furthermore, one of these three parameters/numbers can 
be equal zero, but the three principal conditions for these parameters $\alpha + \beta > 0, \alpha + \gamma > 0$ and $\beta + \gamma > 0$ must always be obeyed. If these three 
parameters are complex, then these three conditions are written in the form $Re(\alpha + \beta) > 0, Re(\alpha + \gamma) > 0$ and $Re(\beta + \gamma) > 0$. These conditions are 
needed to quarantee the convergence of the three-body integrals $\Gamma_{k;l;n}(\alpha, \beta, \gamma)$, Eq.(\ref{int2}).

Now, the explicit formulas for the analogous three-body integrals which 
contain the Bessel functions $j_{0}(k r_2)$ and/or $j_{1}(k r_2)$ is derived 
in the following way. First, consider the integral, Eq.(\ref{int1}), with 
the $j_{0}(x)$ function which can also be written in the form $j_{0}(x) = 
\frac{\sin x}{x}$. In this case the integral Eq.(\ref{int1}) is written as 
the sum of the two following (complex) integrals
\begin{eqnarray}
 I_0(n,\ell; k) = \frac{1}{2 \imath k} \int \int \int \phi_{n\ell}(r_{32}) \exp[-\alpha r_{32} -(\beta - \imath k) r_{31} - \gamma r_{21}] r_{32} r_{21} dr_{32} dr_{31} 
 dr_{21} \label{int5} \\
 - \frac{1}{2 \imath k} \int \int \int \phi_{n\ell}(r_{32}) \exp[-\alpha r_{32} -(\beta + \imath k) r_{31} - \gamma r_{21}] r_{32} r_{21} dr_{32} dr_{31} dr_{21}
 \nonumber
\end{eqnarray}
where $\imath$ is the imaginary unit. Consider the case when the final hydrogen atom is formed in its ground state (or (10)-state). The integral $I_0(n,\ell;k) = I_0(1,0;k)$ can 
easily be determined in perimetric coordinates. Indeed, in this case $\phi_{n\ell}(r_{32}) = \phi_{10}(r_{32}) = 2 \exp(-A r_{32})$, where $A = 1$, and the last integral is reduced to 
the form
\begin{eqnarray}
 I_0(1,0;k) = \frac{1}{\imath k} \int \int \int \exp[-(\alpha + A) r_{32} -(\beta - \imath k) r_{31} - \gamma r_{21}] r_{32} r_{21} dr_{32} dr_{31} dr_{21}
 \label{int51} \\
 - \frac{1}{\imath k} \int \int \int \exp[-(\alpha + A) r_{32} -(\beta + \imath k) r_{31} - \gamma r_{21}] r_{32} r_{21} dr_{32} dr_{31} dr_{21}\nonumber \\
 = - \frac{2 \imath}{k} \int^{\infty}_0 \int^{\infty}_0 \int^{\infty}_0
 \exp[-X_1 u_1 - X_2 u_2 - X_3 u_3] (u_1 + u_3) (u_1 + u_2)
 du_1 du_2 du_3 \nonumber \\
 + \frac{2 \imath}{k} \int^{\infty}_0 \int^{\infty}_0 \int^{\infty}_0
 \exp[-X_1 u_1 - X^{*}_2 u_2 - X^{*}_3 u_3] (u_1 + u_3) (u_1 + u_2)
 du_1 du_2 du_3 \nonumber
\end{eqnarray}
where $X_1 = \alpha + \gamma + A, X_2 = \beta + \gamma - \imath k, X_3 =
\alpha + \beta + A - \imath k$ and $X^{*}_2 = \beta + \gamma + \imath k,
X^{*}_3 = \alpha + \beta + A + \imath k$. Note that each of the last two
integrals is written in the form
\begin{eqnarray}
 J(a,b,c) &=& \int^{\infty}_0 \int^{\infty}_0 \int^{\infty}_0 \exp[-a x - b y - c z] (x + z) (x + y) dz dy dz \nonumber \\
 &=& \frac{1}{a b c} [\frac{2}{a^2} + \frac{1}{a b} + \frac{1}{a c} + \frac{1}{b c}] \label{int52}
\end{eqnarray}
This integral is often called the Coulomb three-body integral, since all matrix elements
of the Coulomb interparticle potential are reduced to this integral. 

The closed analytical expression for this integral allows one to obtain analytical formulas for all 
integrals, Eq.(\ref{int1}), which include the $\mid n \ell \rangle$ radial functions of the final 
hydrogen atom. In reality, such integrals are computed as the partial derivatives from Eq.(\ref{int52}) 
in respect with the parameter $\alpha$. This solves the problem of analytical/numerical calculations of 
the three-body integrals from the exponential functions of the three relative coordinates, 
Eq.(\ref{int1}), which include the Bessel function $j_{0}(k r)$.

Analytical formulas for the analogous integrlas with the Bessel function $j_{1}(k r)$ can be derived 
from Eq.(\ref{int5}) - (\ref{int51}) with the use of the Rayligh relation 
\begin{equation}
  j_1(k r) = - \frac{d j_{0}(k r)}{r dk} \label{bessel}
\end{equation}
known for these two Bessel functions. Analytical and/or numerical computation of the corresponding 
derivatives and integrals is straightforward. In general, all analytical transformations of formulas are 
performed with the use of different platforms for analytical computations such as Mathematica and/or 
Maple (see, e.g., \cite{Wolf} and \cite{Mapl}). Note also that all integrals arising during this procedure 
are regular, since each derivative in respect to $k$ gives an extra power of $r_2$ which compensates the 
presence of this variable in the denominator of the right hand side of Eq.(\ref{bessel}).  

An alternative approach which is also used for calculation of the integrals Eq.(\ref{int1}) is based on our 
formulas derived for the three-particle integrals with the Bessel functions \cite{Fro2013}. The formulas for 
the integrals which contain the Bessel functions $j_0(k r_{32})$ and $j_1(k r_{32})$ are derived in the 
following way. First, let us obtain the computational formula for the following integral 
\begin{eqnarray}
 B^{(0)}_{k;l;n}(\alpha, \beta, \gamma; k) &=& \int_{0}^{+\infty} \int_{0}^{+\infty}
 \int^{r_{32} + r_{31}}_{\mid r_{32} - r_{31} \mid} 
  r^{k}_{32} r^{l}_{31} r^{n}_{21} j_0(k r_{32}) \times \nonumber \\ 
 & & \exp(-\alpha r_{32} - \beta r_{31} - \gamma r_{21}) dr_{32} dr_{31} dr_{21} \label{e33}
\end{eqnarray}
By using the formula $j_0(x) = \frac{\sin x}{x}$ one finds
\begin{eqnarray}
 B^{(0)}_{k;l;n}(\alpha, \beta, \gamma; k) = \sum^{\infty}_{q=0} \frac{(-1)^q k^{2 q}}{(2 q + 1)!} 
 \Gamma_{k + 2 q;l;n}(\alpha, \beta, \gamma) \approx \sum^{q_{max}}_{q=0} \frac{(-1)^q 
 k^{2 q}}{(2 q + 1)!} \Gamma_{k + 2 q;l;n}(\alpha, \beta, \gamma) \label{e34}
\end{eqnarray}
where $\Gamma_{k;l;n}(\alpha, \beta, \gamma)$ is the integral defined in Eq.(\ref{int3}).

The integral $B^{(0)}_{k;l;n}(\alpha, \beta, \gamma; k)$ in the last equation converges for all 
$k$, but for $k \le 1$ it converges very rapidly. In reality, the maximal value of the index $q$ 
(or $q_{max}$) in Eq.(\ref{e34}) is finite. Numerical investigations indicate that to stabilize 15 
decimal digits for $k \le 1$ one needs to use in Eq.(\ref{e34}) $q_{max}$ = 20 to 40. For $k \ge 
2$ the value of $q_{max}$ rapidly increases up to 50 - 70 and even 100. The same conclusion is 
true about the convergence of the three-body integrals with the Bessel function $j_1(x) = 
\frac{\sin x}{x^2} - \frac{\cos x}{x}$. This integral takes the form
\begin{eqnarray}
 B^{(1)}_{k;l;n}(\alpha, \beta, \gamma; k) &=& \int_{0}^{+\infty} \int_{0}^{+\infty} 
 \int^{r_{32} + r_{31}}_{\mid r_{32} - r_{31} \mid} r^{k}_{32} r^{l}_{31} r^{n}_{21} j_{1}(k r_{32}) 
 \exp(-\alpha r_{32} - \beta r_{31} - \gamma r_{21}) dr_{32} dr_{31} dr_{21} \nonumber \\
 &=& \sum^{\infty}_{q=0} \frac{(-1)^q (2 q + 2) 
 k^{2 q + 1}}{(2 q + 3)!} \Gamma_{k + 2 q + 1;l;n}(\alpha, \beta, \gamma) \label{e35}
\end{eqnarray}
The three-body integrals with the lowest order Bessel functions $j_0(x)$ and $j_1(x)$ allows one to determine
all integrals, Eq.(\ref{int1}), needed in calculations of the photodetachment cross-section of the negatively 
charged hydrogen ion. In particular, the integrals $B^{(1)}_{k;l;n}(\alpha, \beta, \gamma; k)$ are needed to 
calculate the photodetachment cross-sections in the $sp-$channel, while the integrals $B^{(0)}_{k;l;n}(\alpha, 
\beta, \gamma; k)$ are needed to evaluate analogous cross-sections in the $ps-$channel. 

To conclude this Section let us present the corresponding integrals with the Bessel functions and a few radial
hydrogenic functions $R_{10}(r), R_{20}(r), R_{21}(r), R_{31}(r)$, where 
\begin{eqnarray}
  R_{10}(r) &=& 2 Q \sqrt{Q} exp(-Q r) \; \; \; , \; \; \; R_{20}(r) = \frac{Q \sqrt{Q}}{\sqrt{2}} \Bigl( 1 
 - \frac12 Q r \Bigr) exp(-\frac{Q r}{2}) \nonumber \\
 R_{21}(r) &=& \frac{Q \sqrt{Q}}{\sqrt{24}} Q r exp(-\frac{Q r}{2}) \; \; \; , \; \; \; 
 R_{31}(r) = \frac{8 Q \sqrt{Q}}{27 \sqrt{6}} \Bigl( 1 - \frac16 Q r \Bigr) exp(-\frac{Q r}{3}) \nonumber
\end{eqnarray}
where $Q = 1$ for the hyrogen atom. For the photodetachment of the hydrogen negative ion only the integrals which 
contain the following products $R_{10}(r_{31}) j_{1}(k r_{32}), R_{20}(r_{31}) j_{1}(k r_{32})$ and $R_{21}(r_{31}) 
j_{0}(k r_{32}), R_{31}(r_{31}) j_{1}(k r_{32})$ are important. In respect with Eqs.(\ref{e34}) and (\ref{e35}) 
these integrals are
\begin{eqnarray}
  & & 2 Q \sqrt{Q} B^{(1)}_{1;1;1}(\alpha, \beta + Q, \gamma; k) \; \; \; , \\  
  & & \frac{Q \sqrt{Q}}{2} B^{(1)}_{1;1;1}(\alpha, \beta + \frac{Q}{2}, \gamma; k) - 
  \frac{Q \sqrt{Q}}{4} B^{(1)}_{1;2;1}(\alpha, \beta + \frac{Q}{2}, \gamma; k)
\end{eqnarray}
and 
\begin{eqnarray}
 & & \frac{Q^2 \sqrt{Q}}{24}  B^{(0)}_{1;2;1}(\alpha, \beta + \frac{Q}{2}, \gamma; k) \; \; \; , \\
 & & \frac{8 Q \sqrt{Q}}{27 \sqrt{6}} B^{(0)}_{1;1;1}(\alpha, \beta + \frac{Q}{3}, \gamma; k) - 
  \frac{4 Q^2 \sqrt{Q}}{81 \sqrt{6}} B^{(0)}_{1;2;1}(\alpha, \beta + \frac{Q}{3}, \gamma; k) \; \; \; ,
\end{eqnarray}
respectively. Analogous formulas can be written for the integrals which include other radial hydrogenic functions, 
e.g., $R_{n0}(r)$ and $R_{m1}(r)$, where $n \ge 3$ and $m \ge 4$. However, the overall contribution of such integrals 
to the photodetachemnt cross-section of the H$^{-}$ ion ($Q = 1$ in all formulas above) is very small and they can be 
neglected in the lowest order approximation.

\section{Simple method to evaluate the photodetachment cross-section}

The photodetachment cross-section of the negatively charged H$^{-}$ ion in the sp-channel can be evaluated with the use of the 
following simple method. The negatively charged hydrogen ion H$^{-}$ is the bound two-electron atomic ion. Let us designate the 
total number of bound electrons in atomic system by $N_e$, while the notation $Q$ stands below for the nuclear charge $Q$. As is 
well known (see, e.g., \cite{BhaDra} - \cite{FroSm03} and references therein) the long range one-electron asymptotic 
of the $N_e-$electron atomic wave 
function is written in the form
\begin{equation}
 \mid \Psi(r) \mid = C r^{\frac{Z}{t} - 1} \exp(-t r) \label{eq32}
\end{equation}
where $C$ is some numerical constant, $Z = Q - N_e + 1$, while $t = \sqrt{2 I_1}$ and $I_1 = \chi_1$ is the first ionization 
potential which corresponds to the dissociation process H$^{-}$ = H + $e^{-}$. For the H$^{-}$ ion one finds in 
Eq.(\ref{eq32}): $Q = 1, N_e = 2$, and therefore, $Z = 0$ and the long range asymptotic of the wave function, Eq.(\ref{eq32}), 
is represented in the Yukawa-type form $\mid \Psi(r) \mid = \frac{C}{r} \exp(-t r)$, where $C$ is a constant which must provide 
the best correspondence with the highly accurate wave function of the H$^{-}$ ion at large $r$. The highly accurate wave 
function of the H$^{-}$ ion is assumed to be known. The explicit formula which is used to determine the constant $C$ is $C = 
\mid \Psi_{{\rm H}^{-}}(r) \mid r \exp(t r)$, where $\Psi_{{\rm H}^{-}}(r) =  \Psi_{{\rm H}^{-}}(r_{32} = r, r_{31} = 0, r_{21} 
= r)$. As follows from the results of actual computations the factor $C$ does not change for relatively large interval of 
variations of $r$, e.g., for $r$ bounded between 7.5 and 35 atomic units. The determined numerical value of this constant $C$ for 
the negatively charged hydrogen ions varies between $\approx 0.12595$ (${}^{\infty}$H$^{-}$) and $\approx 0.12605$ 
(${}^{1}$H$^{-}$).

The photodetachment cross-section $\sigma(H^-)$ of the H$^-$ ion is written in the form \cite{John} (in atomic units)  
\begin{eqnarray}
 \sigma({\rm H}^-) &=& \alpha a^{2}_{0} \frac{p}{2 \pi \omega} \oint \mid {\bf e} \cdot \int \psi^{*}_f \Bigl(\frac{\partial}{\partial {\bf r}}
 \psi_{i}\Bigr) d^3{\bf r} \mid^2 do = \frac{64 \pi^2 C^2 \alpha a^2_0}{3} \cdot \frac{p^3}{(p^2 + \gamma^2)^3} \nonumber \\
 &=& \frac{64 \pi^2 C^2 \alpha}{3} \cdot \frac{p^3}{(p^2 + \gamma^2)^3} \; \; \; .
\end{eqnarray}
where $\alpha$ is the fine structure constant and $a_0$ is the Bohr radius and $\gamma = t = \sqrt{2 I_1}$. The energy conservation law 
(for the photon) is written in the form: $\hbar \omega = \frac{p^2}{2 m_e} + \frac{t^2}{2}$, or (in atomic units): $\omega = \frac{p^2}{2} 
+ \frac{t^2}{2}$. To transofrm the first formula in Eq.(\ref{eq32}) to its final form  we have chosen the incident (one-electron!) wave 
function in the form of Eq.(\ref{eq32}), while the final wave function $\psi_{f}$ is written in the form $exp(\imath {\bf p} \cdot {\bf r})$. 
Also in this equation $p = \mid {\bf p} \mid$ is the momentum of the photo-electron (in atomic units) and $\gamma = 0.23558869473885$ (see 
Table I). In the usual units ($cm^2$) the last formula takes the form
\begin{equation}
 \sigma({\rm H}^-) = 2.7032847 \cdot 10^{-16} \cdot C^2 \cdot \frac{p^3}{(p^2 + \gamma^2)^3} \; \; \; cm^2 \; \; \; .
\end{equation}
This formula with the known value of $C$ can directly be used in calculations of the photodetachment cross-section $\sigma(H^-)$. This
formula essentially coincides with Eq.(8) from \cite{Ohm}. In our case, $C =$ 0.12595, and therefore, from the last formula one finds
\begin{equation}
 \sigma({\rm H}^-) = 4.28840 \cdot 10^{-18} \cdot \frac{p^3}{(p^2 + \gamma^2)^3} \; \; \; cm^2 \; \; \; . \label{simple}
\end{equation}

In real calculations the overall accuracy of this simple formula, Eq.(\ref{simple}), is outstanding. This remakable fact follows, in part, 
from the comparison of the sizes of atomic system $a$ with the wavelengths $\lambda$ of the photons which produce the photodetachment. 
It is clear that $a \approx a_0 \approx 0.529 \AA$ where $a_0$ is the Bohr radius and $a$ is the effective radius of the atomic H$^{-}$ 
ion. On the other hand, the wavelengths $\lambda$ of the photons which produce the photodetachment is $\lambda \approx 1 \cdot 10^5 \AA$. 
This means that $\lambda \gg a$, or, in other words, the ratio $\frac{\lambda}{a} \approx 1 \cdot 10^4$ is a very large number. Briefly, 
we can say that the incident photon sees only an outer asymptotics of the three-body wave function. All details of the wave function of the 
H$^{-}$ ion at small inter-particle distances are not important to describe the photodetachment of this ion. Recently, we have successfully 
used this simple method in analogous applications related to the H$^{-}$, Mu$^{-}$ (or $\mu^{+} e^{-}_2$) and Ps$^{-}$ (or $e^{+} e^{-}_2$) 
ions.

To conclude this Section let us note that some important details of the photodetachment of the H$^{-}$ ion cannot be investigated by applying 
only this simple method. For instance, it cannot be used, in principle, to evaluate the transition probabilities for the $ps-$channel(s) and 
to predict analogous probabilities for the formation of the hydrogen atom in the excited $s-$states in the $sp-$channels. Nevertheless, this
simple method can be useful in astrophysics for fast evaluation of the absorbtion coefficient of the H$^{-}$ ion at different conditions.

\section{Inverse bremsstrahlung in the field of the hydrogen atom}

The knowledge of the photodetachment cross-sections for the $sp-$channles allows one to evaluate the absorbtion of infrared and visible 
radiation by the H$^{-}$ ion in photospheres of many cold stars with surface temperatures $T_s \le$ 8,250 $K$. Addition of the $ps-$channels
increases the overall accuracy of such evaluations. However, at large wavelengths ($\lambda \ge 12,000 \AA$) a different mechanism begin to 
play a significant role in the absorbtion of infrared radiation. A large number of unbound electron which move in the fileds of hydrogen
atoms can absorb a noticeable amount of propagating infrared radiation. In this Section we discuss the electron scattering stimulated by the 
electromagnetic radiation in the field of the heavy hydrogen atom, Eq.(\ref{ee0}). During such a process the radiation energy is transfered, 
in part, to the scattered electrons. Electrons are accelerated, but the mean energy of the radiation field decreases. In classical physics 
this process is called inverse bremsstrahlung. Inverse bremsstrahlung is important in many areas of modern physics and various applications. 
For instance, it is crucially important for workability of explosive devices usually called `hydrogen bombs'. In stellar photoshperes of many 
hydrogen stars inverse bremsstrahlung is responsible for absorption of infrared radiation with $\lambda \ge$ 12,000 $\AA$. This includes our 
Sun \cite{Ohm2}. Furthermore, it is very important to describe the total absorption of infrared radiation in photospheres of many cold stars. 

As mentioned above at small energies of the incident photon $h \nu \le 5 eV$ the original (ground) state of the hydrogen atom does 
not change. This means that for photons of small energies we can ignore all possible excitations of the hydrogen atom, i.e. atomic 
transitions from the ground state of the H atom into other excited states. The incident photon is absorbed exclusively by the 
electron which moves in the field of the neutral hydrogen atom. In this case the electron scattering at the hydrogen atom (H + 
e$^{-}$ + $h \nu$ = H$^{*}$ + e$^{-}$) can be considered as a regular (but non-elastic!) scattering of the electron in the field of the 
heavy hydrogen atom. Note that the hydrogen atom is electrically neutral. In the lowest order approximation we can replace the 
actual hydrogen atom by its polarization potential, which is a central and local potential $V(r) = -\frac{a}{r^{4}}$, where $a$ is 
a positive constant. It is clear that we are dealing with the case of `polarization' inverse bremsstrahlung, which is substantially 
different from the regular (or Coulomb) bremsstrahlung (see, e.g., \cite{Consal}). In particular, for the ${}^{\infty}$H atom in atomic units 
one finds $a = \frac94$. Below, we shall consider the electron scattering in the field of the following central potential
\begin{equation}
   V(r; a, b) = - \frac{a}{(r + b)^{4}} \label{poten}
\end{equation}
which analytically depends upon the two real parameters $a$ and $b$. We need to determine the cross-section of inverse bremsstrahlung 
for an electron which moves in the field of the central potential $V(r; a, b)$, Eq.(\ref{poten}). By using different numerical values 
for the $a$ and $b$ parameters we can obtain a variety of such cross-sections. The main interest, however, will be related to the 
cases when $a \approx 4.5$ and $b \approx 0$, respectively. 

To calculate the cross-section of inverse bremsstrahlung in the field of the neutral hydrogen atom we can use a number of different methods.
Briefly. all such methods can be separated into two large groups. The first group of these methods is based on the use of very approximate 
electronic wave functions which do not describe the electron-electron correlations. By using these wave functions we can solve `exactly' 
the corresponding scattering problem. Moreover, by varying a large number of non-linear parameters included in the incident and final wave
functions one can obtain a good agreement with the known experimental data. The second group of methods is based on an explicit construction 
of the two-electron wave functions. Modern methods allow one to determiner such wave functions to very high numerical accuracy. Hovewer, any
accurate and/or numerical solution of the scattering problem with such wave functions becomes extremely complex and non-accurate. In reality,
we have to face the same old-standing problem here, when the total number of actual variables in the incident and final wave functions is
different from each other. 

In this study we develop an alternative one-electron approach. In our approach the hydrogen atom is replaced by an `effective' central potential. 
This interaction potnetial explicitly depends upon a few parameters whcih can be varied in computations. In the field of such a potential we have 
an infinite number of scattered electronic states. The corresponding one-electron wave functions can be determined to very high accuracy. Finally, 
we have a method which is directly related with the original problem, but it is not overloaded by very complex computational and analytical details. 

Let us describe our method which can be used to calculate cross-section of inverse bremsstrahlung in the field of the hydrogen atom.
First, let us obtain the incident and final wave functions which are the stationary one-electron wave functions of the continuous spectra 
of the radial Schr\"{o}dinger equation (see, e.g., \cite{LLQ}) of the electron moving in the field of the central potential, Eq.(\ref{poten}). 
The solution of one-electron Schr\"{o}dinger equation can be found by representing the unknown wave function $\Psi(r)$ in the form 
$\Psi_{\ell}(r) = A_{\ell}(r) \exp(\imath \delta_{\ell}(r))$. Note that in the field of any central potential the angular momentum $\ell$ is 
a good (i.e. conserving) quantum number. It reduces the original problem to the solution of the following differential equation
\begin{equation}
 \frac{d \delta_{\ell}(r)}{dr} = - \frac{2 V(r; a, b)}{k} \Bigl[ j_{\ell}(k r) \cos \delta_{\ell}(r) - n_{\ell}(k r) \sin \delta_{\ell}(r)
 \Bigr]^2 \label{perphas}
\end{equation}
where the phase $\delta_{\ell}(r)$ of the wave function is assumed to be explicitly $r-$dependent and $\delta_{\ell}(r = 0) = 0$ (initial 
condition). With this initial condition the last equation is reduced to the form of the following integral equation 
\begin{equation}
 \delta_{\ell}(r) = - \frac{2}{k} \int_{0}^{+\infty} V(r; a, b) \Bigl[ j_{\ell}(k r) \cos \delta_{\ell}(r) - n_{\ell}(k r) \sin \delta_{\ell}(r)
 \Bigr]^2 dr \label{phas1}
\end{equation}
The equations Eq.(\ref{perphas}) and Eq.(\ref{phas1}) are the main equations 
of the variable phase method. The method of variable phase has been developed 
by V.V. Babikov and others in the middle of 1960's \cite{Babik} (see also 
references therein). Each of these two equations Eqs.(\ref{perphas}) - 
(\ref{phas1}) can be solved to high accuracy by using various numerical 
methods. In general, if we know the phase function $\delta_{\ell}(r)$ for 
each $r$, then it is possible to determine the amplitude $A_{\ell}(r)$ of the 
wave function and the wave function $\Psi_k(r) = A_{\ell}(r) \exp(\imath 
\delta_{\ell}(r))$ itself. The corresponding equation for the amplitude 
$A_{\ell}(r)$ takes the form \cite{Babik}
\begin{eqnarray}
 \frac{d A_{\ell}(r)}{dr} = - \frac{A_{\ell}(r) V(r; a, b)}{k} \Bigl[ j_{\ell}(k r) \cos \delta_{\ell}(r) - n_{\ell}(k r) \sin \delta_{\ell}(r)
 \Bigr] \Bigl[ j_{\ell}(k r) \sin \delta_{\ell}(r) \nonumber \\ 
 + n_{\ell}(k r) \cos \delta_{\ell}(r) \Bigr] \label{amplit}
\end{eqnarray}
Note that the variable phase method works very well for central potentials which rapidly vanish at $r 
\rightarrow +\infty$, e.g., for $\mid V(r) \mid \sim r^{-n}$, where $n \ge 3$ at $r \rightarrow +\infty$. 
This means that we can apply this method to determine the accurate wave functions of the continuous spectra 
in the field of the polarization potential, Eq.(\ref{poten}), created by the neutral H atom which is 
assumed to be infinitely heavy. It appears that the variable phase method provides a large number of 
advantages in applications to the problems related with the absorption/emission of radiation from the states 
of continuous spectra.

In the case of inverse bremsstrahlung the energy conservation law is written in the form 
\begin{equation}
  \frac{\hbar k^{2}_{i}}{2 m_e} + \omega = \frac{\hbar k^{2}_{f}}{2 m_e} \; \; \; , \; \; or \; \; \;  \hbar \omega = 
  \frac{p^{2}_{f}}{2 m_e} - \frac{p^{2}_{i}}{2 m_e} \label{energ}
\end{equation}
where $k_i$ and $k_f$ are the wave numbers defined as $k_a = \mid {\bf k}_a \mid = \frac{\mid {\bf p}_a \mid}{\hbar}$ and ${\bf p}_a$ is the 
momentum of the particle $a$ and ${\bf k}_a$ is the correspponding wave vector. In atomic units ${\bf k}_a = {\bf p}_a$ and $k_a = p_a$, while
Eq.(\ref{energ}) takes the form $\omega = \frac12 (p^{2}_{f} - p^{2}_{i})$. Also, from Eq.(\ref{energ}) one finds $k_f = \sqrt{k^{2}_{i} + 
\frac{2 m_e}{\hbar} \omega}$. By using the variable phase method described above we can obtain the initial and final state wave functions, i.e. 
$\Psi_{k_i}(r)$ and $\Psi_{k_f}(r)$, respectively. In the variable phase method these wave functions are represented as the products of their 
amplitudes and phase parts, i.e. $\Psi_{\ell;k_i}(r) = A_{\ell;k_i}(r) \exp(\imath \delta_{\ell;k_i}(r))$ and $\Psi_{\ell;k_f}(r) = 
B_{\ell;k_f}(r) \exp(\imath g_{\ell;k_f}(r))$. Let us assume that by using the procedure described above we have determined the incident and 
final wave functions. Now, these two wave functions can be used in the following computations of the cross-section of inverse bremsstrahlung. 

The explicit formula for the cross-section of inverse bremsstrahlung is (in relativistic units)
\begin{equation}
  d\sigma_{{\bf k}{\bf p}_f} = \frac{\omega e^2}{(2 \pi)^4 m_e p_i} \mid {\bf e} \cdot {\bf p}_{i \rightarrow f} \mid^2 do_{{\bf k}} d^3{\bf p}_f  
 \label{formula}
\end{equation}
where $p_i = \mid {\bf p}_i \mid$ is the `absolute' momentum of the incident electron, ${\bf p}_f$ is the momentum of the final electron and 
${\bf p}_{i \rightarrow f}$ is the following matrix element
\begin{equation}
  {\bf p}_{i \rightarrow f} = - \imath \int \Psi^{*}_{k_f} \nabla \Psi_{k_i} d^3{\bf r} \label{moment}
\end{equation}
where $\Psi_{k_i}$ and $\Psi_{k_f}$ are the incident and final one-electron wave functions (see above). Each of these two functions is assumed 
to be properly symmetrized upon all electron-electron permutations. By integrating Eq.(\ref{formula}) over all directions of the final momentum one 
finds the following formula for the cross-section (in atomic units):
\begin{equation}
  d\sigma_{{\bf k} p_f} = \alpha^3 a^{2}_{0} \frac{p^{2}_{f} - p^{2}_{i}}{(2 \pi)^3 p_i} \mid {\bf e} \cdot {\bf p}_{i \rightarrow f} \mid^2 
  do_{{\bf k}} p^2_f d p_f \label{form1}
\end{equation}
After a few steps of straightforward transformations one finds from this formula
\begin{equation}
  d\sigma_{p_i p_f} = \frac{2 \alpha^3 a^{2}_{0}}{3 \pi} \frac{(p^{2}_{f} - p^{2}_{i}) p^2_f}{p_i} \mid \int^{\infty}_0 \Psi_f 
  \Bigl(\frac{d \Psi_{i}}{dr} \Bigr) dr \mid^2 d p_f \label{formula2}
\end{equation}
By using Eqs.(\ref{perphas}) and (\ref{amplit}) we obtain the following expression for the derivative of the $\Psi_i$ function:
\begin{eqnarray}
 \frac{d \Psi_{i}}{dr} &=& \frac{d A_{\ell;k_i}(r)}{dr} \exp(\imath \delta_{\ell;k_i}) +  \imath A_{\ell;k_i}(r) \exp(\imath \delta_{\ell;k_i})
 \frac{d \delta_{\ell;k_i}(r)}{dr} \label{eqphase} \\
 &=& - \frac{V(r; a, b) A_{\ell;k}(r) \exp(\imath \delta_{\ell;k_i}) }{k} \Bigl[ j_{\ell}(k r) \cos \delta_{\ell}(r) - n_{\ell}(k r) \sin \delta_{\ell}(r)
 \Bigr] \Bigl[ j_{\ell}(k r) \sin \delta_{\ell}(r)  \nonumber \\
 &+& n_{\ell}(k r) \cos \delta_{\ell}(r) \Bigr] - \frac{2 V(r; a, b)}{k} A_{\ell;k_i}(r) \exp(\imath \delta_{\ell;k_i}) \Bigl[ j_{\ell}(k r) \cos 
 \delta_{\ell}(r) - n_{\ell}(k r) \sin \delta_{\ell}(r) \Bigr]^2 \nonumber
\end{eqnarray}
or, in other words,
\begin{eqnarray}
 \frac{d \Psi_{i}}{dr} &=& - \frac{V(r; a, b)}{k} \Psi_{i} \Bigl[ j_{\ell}(k r) \cos \delta_{\ell}(r) - n_{\ell}(k r) \sin \delta_{\ell}(r)
 \Bigr] \Bigl[ j_{\ell}(k r) \sin \delta_{\ell}(r) + n_{\ell}(k r) \cos \delta_{\ell}(r) \Bigr] \nonumber \\
 &-& \frac{2 V(r; a, b)}{k} \Psi_{i} \Bigl[ j_{\ell}(k r) \cos \delta_{\ell}(r) - n_{\ell}(k r) \sin \delta_{\ell}(r) \Bigr]^2 \nonumber
\end{eqnarray}
The last equation does not include explicitly the amplitude function $A_{\ell;k_i}(r)$. Note also that the last equation can be re-written in a number
of different forms which are more convenient in some cases. Here we do not wish to discuss this interesting problem. 

Another simplification follows from the fact that the wavelength $\lambda$ of the light quantum which produces photodetachment of the H$^{-}$ ion is 
$\ge$ 12,000 $\AA$, while the effective `scattering zone' $R_s$ for the process, Eq.(\ref{ee0}) is less than 15 - 25 $\AA$. Therefore, we can apply the 
approximation based on the fact that the dimensionless parameter $\frac{R_s}{\lambda} \ll 1$. This approximation allows one to replace the 
$r-$dependent phase factor $\delta_{\ell;k_i}(r)$ in the wave function by its asymptotic value $\delta_{\ell;k_i}(+\infty)$ determined at the 
infinity. In other words, the phase of the wave function does not depend upon $r$, i.e. it is a constant. The wave function $\Psi_{\ell;k_i}(r)$ is 
represented in the form $\Psi_{\ell;k_i}(r) = A_{\ell;k_i}(r) \exp(\imath \delta_{\ell;k_i})$, where $\delta_{\ell} = \delta_{\ell}(r = +\infty)$ is a 
constant (or almost a constant). This drastically simplifies numerical computations of all integrals needed in Eq.(\ref{formula2}). 

Finally, it should be noted that based on the approach developed here we can evaluate the total cross-section of inverse bremsstrahlung for electrons 
scattered by the neutral hydrogen atoms in stellar photoshperes. Currently, the progress in this important area of human knowledge is restricted only by 
the lack of funding.

\section{Conclusion}

We have discussed the absorption of infrared and visible radiation by the negatively charged hydrogen ions H$^{-}$. Such an absorption of radiation 
plays an extremely important role in Solar and stellar astrophysics. In general, absorption of infrared and visible radiation by the H$^{-}$ 
ions is mainly related with the photodetachment of these ions. The absorption of radiation by electrons scattered in the field of the neutral hydrogen 
atom(s) also contributes to the radiation absorbtion coefficient at very large wavelength $\lambda \ge 12,000 \AA$ (infrared radiation). The 
photodetachment cross-section of the H$^{-}$ ion has been determined with the use of highly accurate variational wave functions constructed for 
this ion. We explicitly discuss all details important in calculations of the photodetachment cross-section with the use of highly accurate (or truly 
correlated) wave functions of the two-electron (or three-body) H$^{-}$ ion. To evaluate the total cross-section of electron scattering in the field of 
the polarization potential of the neutral H atom we propose to apply the variable phase method. 

As we mentioned above all additional details of our methods and computational results will be discussed in our next study \cite{Fro2014}. Originally, my plan 
was to combine general theory and numerical results in one paper, but later it was found that the volume of such a manuscript exceeds 50 pages (and many important 
details are still missing). Furthermore, by performing actual calculations I found that some formulas derived for our method(s) are very good and compact only in 
theory. Actual calculations with the use of these formulas lead to slow convergence, numerical instabilities and other problems. For instance, after numerous trials
and corrections I have replaced the final state wave functions of the `free' electron by a different system of functions which is complete and more convenient
in numerical applications. This will be discussed in our next study \cite{Fro2014}. After a few such simple modifications the agreement between our photodetachment 
cross-section and results obtained in earlier works for different wavelength can be considered as very good, but our method is substantially more accurate and 
complete.

\begin{center}
  {\Large Appendix}
\end{center}

In this Appendix we want to test our variational wave functions of the negatively charged hydrogen ${}^{\infty}$H$^{-}, {}^{3}$H$^{-}, {}^{2}$H$^{-}$ 
and ${}^{1}$H$^{-}$ ions in actual computations of the lowest order QED corrections for these two-electron systems. The lowest order QED corrections for 
the ground $1^1S-$state in the negatively charged hydrogen ${}^{\infty}$H$^{-}$ ion is determined by the formula (in atomic units) 
\begin{eqnarray}
 \Delta E^{QED} &=& \frac{8}{3} Q \alpha^3 \Bigl[ \frac{19}{30} - 2 \ln \alpha - \ln K_0 \Bigr] \langle \delta({\bf r}_{eN}) \rangle 
 + \alpha^3 \Bigl[ \frac{164}{15} + \frac{14}{3} \ln \alpha - \frac{10}{3} S(S + 1) \Bigr] \langle \delta({\bf r}_{ee}) \rangle \nonumber \\
 &-& \frac{7}{6 \pi} \alpha^3 \langle \frac{1}{r^{3}_{ee}} \rangle \label{LQED}
\end{eqnarray}
where $\alpha$ is the fine structure constant, $Q$ is the nuclear charge (in atomic units) and $S$ is the total electron spin. For the ground states in 
all H$^{-}$ ions considered in this study we have $S = 0$ (singlet states) in Eq.(\ref{LQED}). The notation $\ln K_0$ in this formula stands for the Bethe 
logarithm \cite{BS}. To simplify numerical calculations in this Appendix one can take the value of $\ln K_0$ from Table 3 in \cite{Fro2005}. The paper  
\cite{Fro2005} also contains references to earlier works in which the formula, Eq.(\ref{LQED}), and all following expressions were derived. 

For the two-electron ions with finite nuclear mass we need to evaluate the corresponding recoil correction to the lowest-order QED correction. Such a 
correction is calculated with the use of the formula (in atomic units) 
\begin{eqnarray}
 \Delta E^{QED}_{M} &=& \Delta E^{QED}_{\infty} - \Bigl(\frac{2}{M} + \frac{1}{M + 1} \Bigr) \Delta E^{QED}_{\infty} + \frac{4 \alpha^3}{3 M} 
 \Bigl[ \frac{37}{3} - \ln \alpha - 4 \ln K_0 \Bigr] \langle \delta({\bf r}_{eN}) \rangle \nonumber \\
 &+& \frac{7 \alpha^3}{3 \pi M} \langle \frac{1}{r^{3}_{eN}} \rangle \label{LQEDA}
\end{eqnarray} 
where $M \gg m_e$ is the nuclear mass. All expectation values in this equation must be determined for the real two-electron ions (with the finite nuclear 
masses). The inverse mass $\frac{1}{M}$ is a small dimensionless parameter which for the ions considered in this study  is smaller than $\le 5 \cdot 
10^{-4}$. The last terms in both Eq.(\ref{LQED}) and Eq.(\ref{LQEDA}) are usually called the Araki-Sucher terms, or Araki-Sucher corrections, since this 
correction was obtained and investigated for the first time in papers by Araki and Sucher. The expectation value of the term $\langle \frac{1}{r^{3}_{ee}} 
\rangle$ is singular, i.e. it contains the regular and non-zero divergent parts. General theory of the singular exponential integrals was developed in our 
earlier works (see, e.g., \cite{HFS} and \cite{Fro2005}). In particular, in \cite{Fro2005} we have shown that the $\langle \frac{1}{r^{3}_{ee}} \rangle$ 
expectation value is determined from the following formula
\begin{eqnarray}
  \langle \frac{1}{r^{3}_{ee}} \rangle = \langle \frac{1}{r^{3}_{ee}} \rangle_R + 4 \pi \langle \delta({\bf r}_{ee}) \rangle
\end{eqnarray} 
where $\langle \frac{1}{r^{3}_{ee}} \rangle_R$ is the regular part of this expectation value and $\langle \delta({\bf r}_{ee}) \rangle$ is the expectation 
value of the electron-electron delta-function. Briefly, we can say that the overall contribution of the signular part of the $\frac{1}{r^{3}_{ee}}$ operator 
is reduced to the expectation value of the corresponding delta-function.

It follows from the formulas Eqs.(\ref{LQED}) and (\ref{LQEDA}) that highly accurate evaluation of the lowest-order QED correction is a very effective test
of the non-relativistic wave functions which can be later used in numerical calculations of the photodetachment cross-section(s) and absorption coefficients 
of the negatively charged hydrogen ions. Indeed, the formulas Eqs.(\ref{LQED}) and (\ref{LQEDA}) contain the electron-nuclear and electron-electron 
delta-functions. To prove the correctness of these values and approximately evaluate the total number of stable decimal digits in each of the delta-functions
one need to compute the corresponding cusp values and compare them with the expected (or predicted) values. This essentially coincides with the standard test
for the non-relativistic wave functions discussed in the main text. In addition to these delta-functions Eqs.(\ref{LQED}) and (\ref{LQEDA}) include the 
$\langle \frac{1}{r^{3}_{ij}} \rangle$ expectation value which is singular. Accurate numerical evaluation of the $\langle \frac{1}{r^{3}_{eN}} \rangle$ and
$\langle \frac{1}{r^{3}_{ee}} \rangle$ expectation values is not an easy task (see, e.g., \cite{Fro2005}), but it provides another effective test for the 
non-relativistic wave function. Our results for the lowest-order QED correction for the ${}^{\infty}$H$^{-}, {}^{3}$H$^{-}, {}^{2}$H$^{-}$ and ${}^{1}$H$^{-}$ 
ions can be found in Table V (in atomic units). The lowest-order QED corrections $\Delta E^{QED}$ and $\Delta E^{QED}_{M}$ determined for each of these ion 
with the finite nuclear mass are the most accurate values to-date. In reality, they coincide in seven first decimal digits with the corresponding values from 
\cite{Fro2005}, where we used less accurate variational wave functions. Very likely, the lowest-order QED corrections are of interest by themselves, but we 
have calculated them here only to prove the high quality of our non-relativistic wave functions.

\newpage
\begin{table}[tbp]
   \caption{The non-relativistic total energies $E$ (in $a.u.$) determined with different number 
            of basis functions $N$, the asymptotic values of the total energies $E(N = \infty)$ 
            (in $a.u.$) and ionization/detachment potential $\chi_1$ (in $eV$) of the hydrogen 
            negatively charged ions (isotopes) in atomic units.}
     \begin{center}
     \begin{tabular}{| c | c | c | c | c |}
      \hline\hline
 $K$ & $E$(${}^{\infty}$H$^{-}$) & & &  $E$(${}^{3}$H$^{-}$) \\
     \hline
 3500 & -0.527751 016544 377196 590213 & & & -0.527649 048201 920733 538766 \\
 
 3700 & -0.527751 016544 377196 590333 & & & -0.527649 048201 920733 538885 \\

 3840 & -0.527751 016544 377196 590389 & & & -0.527649 048201 920733 538983 \\

 4000 & -0.527751 016544 377196 590446 & & & -0.527649 048201 920733 539050 \\
     \hline
 $E(N = \infty)$ & -0.527751 016544 377196 59075(10) & & & -0.527649 048201 920733 539355(10) \\
     \hline
 $\chi_1$ & -0.755143 903366 701124 238 & & & -0.754843 900893 517142 964 \\
        \hline\hline

 $K$ & $E$(${}^{2}$H$^{-}$) & & &  $E$(${}^{1}$H$^{-}$) \\

        \hline
 3500 & -0.527598 324689 706528 595827 & & & -0.527445 881119 767477 071142 \\

 3700 & -0.527598 324689 706528 595960 & & & -0.527445 881119 767477 071261 \\

 3840 & -0.527598 324689 706528 596044 & & & -0.527445 881119 767477 071359 \\

 4000 & -0.527598 324689 706528 596110 & & & -0.527445 881119 767477 071425 \\
     \hline
$E(N = \infty)$ & -0.527598 324689 706528 596355(10) & & & -0.527445 881119 767477 071665(10) \\
     \hline
  $\chi_1$ & -0.754694 721951 430623 253 & & & -0.754246 603605 793792 368 \\
    \hline \hline
  \end{tabular}
  \end{center}
  \end{table}
%


\begin{table}[tbp]
   \caption{Convergence of the expectation values of some electron-nuclear and electron-electron properties for the 
    ${}^{\infty}$H$^{-}$ ion. All values are shown in atomic units and $N$ is the total number of basis functions used.}
     \begin{center}
     \begin{tabular}{| c | c | c | c |}
      \hline\hline
  & $\langle r_{eN} \rangle$ & $\langle r^2_{ee} \rangle$ &  $\langle r^{-2}_{eN} \rangle$ \\ 
     \hline
 3500 & 2.710178278444420365208 & 25.20202529124033189252 & 1.11666282452545228 \\ 
 
 3700 & 2.710178278444420365265 & 25.20202529124033189563 & 1.11666282452544953 \\ 

 3840 & 2.710178278444420365286 & 25.20202529124033189660 & 1.11666282452544421 \\ 

 4000 & 2.710178278444420365301 & 25.20202529124033189738 & 1.11666282452543860 \\ 
        \hline
 & $\tau_{eN}^{(a)}$ & $\langle -\frac12 \nabla^2_e \rangle$ & $\langle {\bf r}_{eN} \cdot {\bf r}_{ee} \rangle$ \\
       \hline
 3500 & 0.6451445424122193894344 & 0.26387550827218859829282 & 12.60101264562016594726 \\

 3700 & 0.6498715811920881669355 & 0.26387550827218859829304 & 12.60101264562016594781 \\

 3840 & 0.6498715811920881669351 & 0.26387550827218859829337 & 12.60101264562016594830 \\

 4000 & 0.6498715811920881669349 & 0.26387550827218859829358 & 12.60101264562016594869 \\
     \hline \hline
  \end{tabular}
  \end{center}
${}^{(a)}$The notation $\tau_{eN}$ stands for the $\langle \cos ({\bf r}_{31} {}^{\wedge} {\bf r}_{21}) \rangle = 
\langle \cos ({\bf r}_{eN} {}^{\wedge} {\bf r}_{ee}) \rangle$ expectation value. Note also, that for all two-electron atomic systems 
$2 \tau_{eN} + \tau_{ee} = 1 + 4 \langle f \rangle$, where $\langle f \rangle$ is the expectation value mentioned in Table III.
  \end{table}


\begin{table}[tbp]
   \caption{The expectation values of some propeties (in atomic units) for the ${}^{\infty}$H$^{-}$ ion.}
     \begin{center}
     \begin{tabular}{| c | c | c | c |}
      \hline\hline
 $\langle r^{-2}_{eN} \rangle$ & $\langle r^{-2}_{ee} \rangle$ & $\langle r^{-1}_{eN} \rangle$ & $\langle r^{-1}_{ee} \rangle$ \\
      \hline 
 1.11666282452542572 & 0.15510415256242466 & 0.6832617676515272224 & 0.311021502214300052 \\
       \hline
 $\langle r_{eN} \rangle$ & $\langle r_{ee} \rangle$ & $\langle r^{2}_{eN} \rangle$ & $\langle r^{2}_{ee} \rangle$ \\
      \hline
 2.7101782784444203653 & 4.4126944979917277211 & 11.913699678051262274 & 25.202025291240331897 \\
      \hline
 $\langle r^{3}_{eN} \rangle$ & $\langle r^{3}_{ee} \rangle$ & $\langle r^{4}_{eN} \rangle$ & $\langle r^{4}_{ee} \rangle$ \\
      \hline
 76.02309704902717911 & 180.60560023017477483 & 645.144542412219375 & 1590.09460393948530 \\
      \hline\hline
 $\langle [r_{32} r_{31}]^{-1} \rangle$ & $\langle [r_{eN} r_{ee}]^{-1} \rangle$ & $\langle [r_{32} r_{31} r_{21}]^{-1} \rangle$ & $\langle \delta({\bf r}_{eeN}) \rangle$ \\
     \hline
 0.38262789034020545 & 0.25307756706456687 & 0.20082343962918944 & 5.129778775490$\cdot 10^{-3}$ \\
     \hline\hline
 $\langle \delta({\bf r}_{eN}) \rangle$ & $\nu_{eN}^{(a)}$ & $\langle \delta({\bf r}_{eN}) \rangle$ & $\nu_{ee}^{(a)}$ \\
      \hline
 0.1645528728473590 & -1.00000000001778 & 2.737992126104611$\cdot 10^{-3}$ & 0.500000002446 \\
     \hline
 $\tau_{eN}$ & $\tau_{ee}$ & $\langle f \rangle$ & $\langle {\bf r}_{31} \cdot {\bf r}_{32} / r^3_{31} \rangle$ \\
      \hline
 0.6498715811920881669 & -0.1051476935659779011 & 0.048648867204549608205 & -0.4642618530806317014 \\
     \hline\hline 
 $\langle -\frac12 \nabla^2_e \rangle$ & $\langle -\frac12 \nabla^2_N \rangle$ & $\langle \nabla_e \cdot \nabla_e \rangle$ & $\langle \nabla_e \cdot \nabla_N \rangle$ \\
      \hline
 0.2638755082721885983 & 0.560630798396681918 & 0.0328797818523047217 & -5.60630798396681918 \\
     \hline
 $\langle {\bf r}_{eN} \cdot {\bf r}_{ee} \rangle$ & $\langle (r^{-3}_{eN})_R \rangle$ & $\langle (r^{-3}_{ee})_R \rangle$ & $\langle r^{-3}_{eN} \rangle$ \\
      \hline
 12.601012645620165948 & -3.43559485054432 & 0.064307887285283 & -1.36776246468689 \\ 
    \hline \hline
  \end{tabular}
  \end{center}
 ${}^{(a)}$The expected cusp values (in $a.u.$) for the ${}^{\infty}$H$^{-}$ ion are $\nu_{eN} = -1.0$ and $\nu_{ee} = 0.5$ (exactly).
  \end{table}
%

\begin{table}[tbp]
   \caption{Convergence of the expectation values of the electron-nuclear and electron-electron delta-functions 
    ($\langle \delta({\bf r}_{eN}) \rangle$ and $\langle \delta({\bf r}_{ee}) \rangle$) and regular parts od the inverse cubic  
    expectation values ($\langle r^{-3}_{eN} \rangle_R$ and $\langle r^{-3}_{ee} \rangle_R$) for the ${}^{3}$H$^{-}, {}^{2}$H$^{-}$ 
    and ${}^{1}$H$^{-}$ ions. All values are shown in atomic units and $N$ is the total number of basis functions used.}
     \begin{center}
     \begin{tabular}{| c | c | c | c |}
      \hline\hline
  & ${}^{3}$H$^{-}$ (tritium) & ${}^{2}$H$^{-}$ (deuterium) & ${}^{1}$H$^{-}$ (protium) \\
     \hline
  & $\langle \delta({\bf r}_{eN}) \rangle$ & $\langle \delta({\bf r}_{eN}) \rangle$ & $\langle \delta({\bf r}_{eN}) \rangle$ \\
     \hline
 3500 & 0.16446163681098 & 0.16441626335994 & 0.16427994412837 \\
 
 3700 & 0.16446163681128 & 0.16441626336024 & 0.16427994412867 \\

 3840 & 0.16446163681069 & 0.16441626335966 & 0.16427994412847 \\

 4000 & 0.16446163681105 & 0.16441626336002 & 0.16427994412844 \\
     \hline
  & $\langle \delta({\bf r}_{ee}) \rangle$ & $\langle \delta({\bf r}_{ee}) \rangle$ & $\langle \delta({\bf r}_{ee}) \rangle$ \\
     \hline
 3500 & 2.735845627669$\cdot 10^{-3}$ & 2.734778360513$\cdot 10^{-3}$ & 2.731572792726$\cdot 10^{-3}$ \\
 
 3700 & 2.735845627652$\cdot 10^{-3}$ & 2.734778360496$\cdot 10^{-3}$ & 2.731572792710$\cdot 10^{-3}$ \\ 

 3840 & 2.735845627712$\cdot 10^{-3}$ & 2.734778360555$\cdot 10^{-3}$ & 2.731572792769$\cdot 10^{-3}$ \\ 

 4000 & 2.735845627776$\cdot 10^{-3}$ & 2.734778360619$\cdot 10^{-3}$ & 2.731572792833$\cdot 10^{-3}$ \\
     \hline
     \hline
  & $\langle r^{-3}_{eN} \rangle_R$ & $\langle r^{-3}_{eN} \rangle_R$ & $\langle r^{-3}_{eN} \rangle_R$ \\
     \hline
 3500 & -3.43332137343396 & -3.43219080091021119 & -3.4287944147910614 \\
 
 3700 & -3.43332137346264 & -3.43219080093883454 & -3.4287944148194641 \\

 3840 & -3.43332137341561 & -3.43219080089185737 & -3.4287944147726374 \\

 4000 &  -3.4333213734460 & -3.43219080092215666 & -3.4287944148027678 \\
     \hline
  & $\langle r^{-3}_{ee} \rangle_R$ & $\langle r^{-3}_{ee} \rangle_R$ & $\langle r^{-3}_{ee} \rangle_R$ \\
     \hline
 3500 & 6.427894892817$\cdot 10^{-2}$ & 6.426455565395$\cdot 10^{-2}$ & 6.422130608689$\cdot 10^{-2}$ \\
 
 3700 & 6.427894893018$\cdot 10^{-2}$ & 6.426455565595$\cdot 10^{-2}$ & 6.422130608885$\cdot 10^{-2}$ \\ 

 3840 & 6.427894892413$\cdot 10^{-2}$ & 6.426455564991$\cdot 10^{-2}$ & 6.422130608281$\cdot 10^{-2}$ \\

 4000 & 6.427894892942$\cdot 10^{-2}$ & 6.426455564520$\cdot 10^{-2}$ & 6.422130608203$\cdot 10^{-2}$ \\
    \hline \hline
  \end{tabular}
  \end{center}
  \end{table}


\begin{table}[tbp]
   \caption{The Bethe logarithm (in $a.u.$) and lowest order QED corrections $E^{QED}$ and $E^{QED}_M$ (in $MHz$) for the ground 
            $1^1S-$state(s) in the ${}^{\infty}$H$^{-}, {}^{3}$H$^{-}, {}^{2}$H$^{-}$ and ${}^{1}$H$^{-}$ ions.}
     \begin{center}
     \begin{tabular}{| c | c | c | c | c |}
      \hline\hline
                    & ${}^{\infty}$H$^{-}$ & ${}^{3}$H$^{-}$ (tritium) & ${}^{2}$H$^{-}$ (deuterium) & ${}^{1}$H$^{-}$ (protium) \\
       \hline 
 $\ln K_0$ ($a.u.$) & 2.993004415 & 2.993011414 & 2.993014897 & 2.993025369 \\
    \hline
 $\Delta E^{QED}$ & 8215.203465 & 8210.661159 & 8208.402174 & 8201.615291 \\
    \hline
 $\Delta E^{QED}_{M}$ & 8215.203465 & 8207.191206 & 8203.207102 & 8191.239514 \\
   \hline \hline
  \end{tabular}
  \end{center}
  \end{table}
\end{document}